\documentclass[twoside]{article}

\usepackage[accepted]{aistats2022}
\usepackage[round]{natbib}

\usepackage[utf8]{inputenc} 
\usepackage[T1]{fontenc}    
\usepackage{hyperref}       
\usepackage{url}            
\usepackage{booktabs}       
\usepackage{amsfonts}       
\usepackage{nicefrac}       
\usepackage{microtype}      
\usepackage{setspace}
\usepackage{amssymb}
\usepackage{url}
\usepackage{lettrine}
\usepackage{framed}
\usepackage{color}
\usepackage{algorithm}
\usepackage{algorithmic}
\usepackage{overpic}
\usepackage{dsfont}
\usepackage{booktabs} 
\usepackage{framed}
\usepackage{bm}
\usepackage{bbm}
\usepackage{ifthen}
\usepackage{graphicx}
\usepackage{dsfont}
\usepackage{framed}
\usepackage{bm}
\usepackage{bbm}
\usepackage{overpic}
\usepackage[svgnames]{xcolor}
\newtheorem{theorem}{Theorem}
\newtheorem{proposition}{Proposition}
\newtheorem{lemma}{Lemma}
\usepackage{xr}
\usepackage[capitalise]{cleveref}
\usepackage{balance}
\def\R{\mathbb{R}}
\def\E{\mathbb{E}}

\def\log{\mathrm{log}}
\def\exp{\mathrm{exp}}

\newcommand{\norm}[1]{\left\lVert#1\right\rVert}

\newcommand{\inv}[1]{\frac{1}{#1}}

\newenvironment{proof}{\paragraph{Proof:}}{\hfill$\square$}

\newcommand{\editr}[1]{{\color{black} #1}}
\newcommand{\algorithmicinit}{\textbf{Initialize:}}
\newcommand{\INIT}{\item[\algorithmicinit]}
\newcommand{\abs}[1]{\left|#1\right|}

\hypersetup{
	colorlinks=true,
	linkcolor=blue,
	urlcolor=red,
	citecolor=DarkBlue,
	linkbordercolor={0 0 1}
}

\begin{document}
%

%

\twocolumn[

\aistatstitle{\texttt{$k$-experts} - Online Policies and Fundamental Limits}

\aistatsauthor{Samrat Mukhopadhyay \And Sourav Sahoo \And  Abhishek Sinha }

\aistatsaddress{ Indian Institute of Technology (ISM)\\ Dhanbad\And  Indian Institute of Technology\\ Madras \And Indian Institute of Technology \\Madras } ]

\begin{abstract}
We introduce the \texttt{$k$-experts} problem - a generalization of the classic \emph{Prediction with Expert's Advice} framework. Unlike the classic version, where the learner selects exactly one expert from a pool of $N$ experts at each round, in this problem, the learner can select a subset of $k$ experts at each round $(1\leq k\leq N)$. The reward obtained by the learner at each round is assumed to be a function of the $k$ selected experts. The primary objective is to design an online learning policy with a small regret. In this pursuit, we propose \texttt{SAGE} (\textbf{Sa}mpled Hed\textbf{ge}) - a framework for designing efficient online learning policies by leveraging statistical sampling techniques. 
For a wide class of reward functions, we show that \texttt{SAGE} either achieves the first sublinear regret guarantee or improves upon the existing ones. 
Furthermore, going beyond the notion of regret, we fully characterize the mistake bounds achievable by online learning policies for stable loss functions. We conclude the paper by establishing a tight regret lower bound for a variant of the \texttt{$k$-experts} problem and carrying out experiments with standard datasets.
 \end{abstract}

\section{INTRODUCTION} \label{intro1}
The classic \emph{Prediction with Expert's Advice} problem, also known as the \texttt{Experts} problem in the literature, is a canonical framework for online learning \citep{cesa2006prediction}. This problem is usually formulated as a two-player sequential game played between a learner and an adversary. Consider a set of $N$ experts indexed by the set $[N]=\{1,2,\ldots, N\}.$ At each round $t$, the adversary secretly selects a reward vector $\bm{r}_{t} \in [0,1]^N$ for the experts. 
At the same time (without knowing the rewards for the present round), the learner selects an expert (possibly randomly) and then receives a reward equal to the reward of the chosen expert. The goal of the learner is to design an online learning policy that incurs a small \emph{regret}. Recall that the regret of an online learning policy over a given time horizon is defined as the difference between the reward accumulated by the best fixed expert in hindsight and the total expected reward accrued by the policy  (see Eqn.\ \eqref{regret-def}). Many online learning policies achieving sublinear regrets in this setting are known, most notably, \texttt{Hedge} \citep{vovk1998game, freund1997decision}. 

In this paper, we initiate the study of the \texttt{$k$-experts} problem - a generalization of the above \texttt{Experts} framework. The \texttt{$k$-experts} problem arises in many settings, including online ad placement, personalized news recommendation, adaptive feature selection, and paging. In the \texttt{$k$-experts} problem, instead of selecting only one expert at each round, the learner selects a subset $S_t \subseteq [N]$ containing $k$ experts at each round $t$ ($1 \leq k \leq N$). The reward $q(S_t)$ received by the learner at round $t$ depends on the rewards of the experts in the chosen set $S_t$. Table \ref{k-experts-variants} lists some variants of the \texttt{$k$-experts} problem considered in this paper.
\begin{table*}[t]
  \caption{Variants of the \texttt{$k$-experts} problem}
  \label{k-experts-variants}
  \centering
  \begin{tabular}{llll}
    \toprule
    \texttt{Sum-reward}     &  \texttt{Max-reward} & \texttt{Pairwise-reward}& \texttt{Monotone reward} \\
    \midrule
    $q_{\texttt{sum}}(S_t) = \sum_{i \in S_t} r_{ti}$ & $q_{\texttt{max}}(S_t) = \max_{i \in S_t} r_{ti}$  & $q_{\texttt{pair}}(S_t) = \sum_{i,j \in S_t} r_{it}r_{jt}$ & $q_{\texttt{monotone}}(S_t)= f_t(S)$    \\
       \bottomrule
  \end{tabular}
\end{table*}
 In the \texttt{Sum-reward} variant, the reward accrued by the learner at round $t$ is given by the sum of the rewards of the experts in the chosen set $S_t$. In particular, let $\bm{p}_{ti}$ denote the (conditional) marginal probability that the $i$\textsuperscript{th} expert is included in the set $S_t$, given the history $\mathcal{F}_{t-1}$ of the game up to round $t-1$. Then, we can express the (conditional) expected reward for the $t$\textsuperscript{th} round as $\mathbb{E}[q_{\texttt{sum}}(S_t)|\mathcal{F}_{t-1}]= \mathbb{E}[\sum_{i\in S_t} r_{ti}|\mathcal{F}_{t-1}] = \langle \bm{r}_t, \bm{p}_t\rangle .$ 
  However, unlike the \texttt{Sum-reward} variant, the expected accrued reward for other variants depends on higher-order joint inclusion probabilities as well (as opposed to only marginals).  
In our most general case, apart from monotonicity, we \emph{do not} impose any other condition (\emph{e.g.,} submodularity \citep{streeter2007online}) on the reward function. 
For each of the above variants, we consider the problem of designing an online expert selection policy that minimizes the regret $\mathcal{R}_T$ (or a variant of it) over a horizon of length $T$:
\begin{eqnarray} \label{regret-def}
	\mathcal{R}_T = \max_{S:|S|=k} \sum_{t=1}^T q(S) - \sum_{t=1}^T \mathbb{E}q(S_t).
\end{eqnarray}
In the above, the expectation in the second term is taken with respect to any randomness of the learner.

%
%
%

 \paragraph{Related work:} A special case of the \texttt{$k$-experts} problem is \emph{Online $N$-ary prediction with} \texttt{$k$-sets}, which we briefly refer to as the \texttt{$k$-sets} problem \citep{koolen2010hedging}. In this problem, a learner sequentially predicts the next symbol for an unknown $N$-ary sequence $\bm{y} = (y_1, y_2, \ldots, y_T)$ chosen by an adversary. The symbols are revealed to the learner sequentially in an online fashion. However, instead of predicting a single symbol $\hat{y}_t \in [N]$ at each round, the learner is allowed to output a subset $S_t,$ consisting of $k$ symbols at round $t$. The learner's prediction for round $t$ is considered to be correct if and only if the predicted set $S_t$ contains the true symbol $y_t$. In the event of a correct prediction, the learner receives unit reward, else, it receives zero rewards for that round. The goal of the learner is to maximize its cumulative reward over a given time horizon. 
 It is easy to see that the above problem is a  special case of the \texttt{$k$-experts} problem with the \texttt{Sum-reward} variant, where the adversary's actions are constrained as $r_{ti} \in \{0,1\}$ with $\sum_{i=1}^N r_{ti}=1, \forall t,i.$ 
 In a seminal paper, \citet{cover1966behavior} studied the fundamental limits of online binary prediction, which is a special case of the \texttt{$k$-sets} problem with $N=2$ and $k=1$. Cover gave a complete characterization of the set of all \emph{stable} reward profiles achievable by online policies (see Section \ref{cover} for the definition of stability). Fifty years later, \citet{rakhlin2016tutorial} generalized Cover's result to an arbitrary alphabet of size $N$, but still requiring $k=1$. The characterization of the prediction error for the \texttt{$k$-sets} problem for an arbitrary $N$ and $k$ has been a long-standing open problem.  

Coming back to the problem of minimizing the static regret for the \texttt{$k$-sets} problem, a quick-and-dirty approach can be used to reduce the problem to an instance of the classic \texttt{Experts} problem with a much larger set of experts, which we call \emph{meta-experts}. In this reduction, a meta-expert is identified with one of the $\binom{N}{k}$ possible subsets of experts of size $k$. One can then use any known low-regret prediction policy, such as \texttt{Hedge}, on the meta-experts to design an online learning policy for the \texttt{$k$-sets} problem.  \citet{koolen2010hedging} referred to the resulting \texttt{Hedge} policy as \texttt{Expanded Hedge}. An obvious challenge with this approach is to overcome the severe computational inefficiency of the resulting online policy, which, \emph{apparently}, needs to keep track of exponentially many experts. To resolve this issue, \cite{koolen2010hedging} proposed the \emph{Component Hedge} (\texttt{CH}) algorithm 
and showed that the proposed policy yields a tight regret bound. However, the \texttt{CH} algorithm involves a projection and decomposition step, each of which costs $O(N^2)$. Although the projection step was later shown to be implementable in linear time {\citep[Theorem 7]{herbster2001tracking}}, the best-known algorithm for the decomposition step still takes $O(N^2)$ time {\cite[Algorithm 2]{warmuth2008randomized}}. The work~{\cite{suehiro2012online}} speculates the existence of an $O(N \log N)$ algorithm for the decomposition. However, their algorithm (Algorithm 4) and its  analysis mentioned in Theorem 10 of the paper still has $O(N^2)$ complexity. We refer the readers to \citet{takimoto2013efficient} for an excellent survey of the efficient projection and decomposition schemes for the \texttt{$k$-sets} and other online combinatorial optimization problems. The \texttt{$k$-sets} problem has also been investigated by \citet{daniely2019competitive}, as an instance of the \emph{paging} problem. 
The authors alleviated the complexity of the naive \texttt{Hedge} implementation by reducing it to a problem of sequential sampling from a recursively defined distribution. Unfortunately, the resulting policy is still sufficiently complex ($\Omega(N^2)$). Recently, \citet{sigmetrics20} studied the paging problem and proposed an efficient and regret-optimal \emph{Follow-the-Perturbed-Leader}-style policy. Although simple to implement, their algorithm does not admit an adaptive regret bound. \editr{Finally, 
the papers \cite{krause2014submodular, streeter2007online, harvey2020improved} studied online maximization of monotone submodular reward functions. However, the problem of achieving sublinear regret for arbitrary monotone reward functions has been wide open.}
\paragraph{Our contributions:}
We make the following contributions in this paper: 
\begin{enumerate}
\item In Section \ref{sage-intro}, we introduce \texttt{SAGE} - an efficient, projection-free, regret-optimal online prediction framework. 

\item In Section \ref{cover}, we  generalize \cite{cover1966behavior}'s result on binary sequence prediction by characterizing the set of all stable error profiles achievable by online learning policies for the $N$-ary prediction problem with $k$-sets.

\item 
 In Section \ref{hedge-k-set}, we design two online policies for the \texttt{$k$-sets} problem using the \texttt{SAGE} framework. The policy in Section \ref{hedge-sec} runs in linear time, admits an adaptive regret bound, and overcomes the existing quadratic computational barrier (see Table \ref{comp-table}). The policy uses standard sub-routines such as Fast Fourier Transform and Madow's sampling. In Section \ref{ftrl}, we propose another prediction policy for the \texttt{$k$-sets} problem based on the \texttt{FTRL} framework. 
 
  
\item In Section \ref{pairwise_section}, using the \texttt{SAGE} framework, we design an \emph{improper} learning policy that achieves $O(\sqrt{T})$ regret for the \texttt{Pairwise-reward} variant of the \texttt{$k$-experts} problem.
 
\item \editr{ In Section \ref{general_reward}, we use the \texttt{SAGE} framework to design an efficient online prediction policy for arbitrary \texttt{Monotone reward} functions. This policy works by approximating the reward with modular functions. To the best of our knowledge, this is the \emph{first} online learning policy for arbitrary monotone reward functions with a guaranteed $O(\sqrt{T})$ regret. }

\item In Section \ref{lower bound}, we establish a tight regret lower bound for the \texttt{Max-reward} version of the \texttt{$k$-experts} problem.

\end{enumerate}
We conclude this section by giving a brief overview and key intuition for the \texttt{SAGE} framework. 
\subsection{Key Insights for \texttt{SAGE}} \label{sage-intro}

We begin our discussion with the \texttt{Sum-reward} variant in the \texttt{$k$-experts} problem. 
 As pointed out earlier, the expected sum reward obtained by any policy depends \emph{only} on the first-order marginal inclusion probabilities and \emph{not} on the higher-order joint distribution. In particular, any two online prediction policies, that have the same conditional marginal inclusion probabilities, yield \emph{exactly} the same reward per round. This simple observation leads to the \texttt{SAGE} meta-algorithm described in Algorithm \ref{sage}. 
\begin{algorithm} 
\caption{The Generic \texttt{SAGE} Meta-Algorithm}
\label{sage}
\begin{algorithmic}[1]
\STATE Start with a low-regret base online prediction policy $\pi_{\textrm{base}}$ (\emph{e.g.,} \texttt{Hedge}). We \textbf{do not} require the base policy $\pi_{\textrm{base}}$ to be computationally efficient.  
\FOR {each round $t$}
\STATE Efficiently compute the first-order marginal inclusion probabilities ($\bm{p}_t$) corresponding to the policy $\pi_{\textrm{base}}.$ This step amounts to marginalizing the joint distribution induced by the policy $\pi_{\textrm{base}}$.  
\STATE Efficiently sample $k$ elements according to the marginal distribution $\bm{p}_t$ computed above.
\ENDFOR
\end{algorithmic}
\end{algorithm}
 From the pseudocode, it is clear that the \texttt{SAGE} meta-algorithm has the same regret as the base policy $\pi_{\textrm{base}}$. However,  
 unlike the base policy (which could be computationally intractable), the \texttt{SAGE} policy can be efficiently implemented in many problems. For example, we show in Section \ref{hedge-sec} that when \texttt{Hedge} is used as the base policy for the \texttt{$k$-sets} problem, the marginalization in line 3 reduces to the evaluation of certain elementary symmetric polynomials. These quantities can be efficiently computed using Fast Fourier Transform techniques. Furthermore, an efficient sampler for line $4$ can be borrowed from the statistical sampling literature, reviewed in Section \ref{prelims}.  
 
Note that \texttt{SAGE} is not necessarily regret-optimal for arbitrary monotone reward functions where the expected reward depends on higher-order inclusion probabilities. However, in Section \ref{general_reward}, we show that we can still use the \texttt{SAGE} framework in this case by approximating the given reward function with modular reward functions. The approximation utilizes recent results from non-submodular set function optimization theory.
\begin{table*}
  \caption{Performance comparison among different policies for the \texttt{$k$-sets} problem}
  \label{comp-table}
  \centering
  \begin{tabular}{llll}
    \toprule
    Policies     &  Reference & Regret bound    & Complexity \\
    \midrule
    \texttt{FTPL} (Gaussian perturbation) & \cite{cohen2015following}  & $2\sqrt{2k^2T\ln \frac{Ne}{k}}$ & $\tilde{O}(N)$     \\
    Component Hedge     & \cite{koolen2010hedging} & $\sqrt{2kT \ln\frac{N}{k}}$& $O(N^2)$      \\
    \texttt{SAGE} (with $\pi_{\textrm{base}}= $\texttt{ Hedge})     & This paper      & $\sqrt{2kT \ln \frac{Ne}{k}}$& $\tilde{O}(N)$  \\
    \texttt{SAGE} (with $\pi_{\textrm{base}}= $\texttt{ FTRL} )     & This paper      & $2\sqrt{2kT \ln \frac{N}{k}}$& $\tilde{O}(N)$  \\
    \bottomrule
  \end{tabular}
\end{table*}

\section{PRELIMINARIES: SAMPLING WITHOUT REPLACEMENT} \label{prelims}
The proposed \texttt{SAGE} meta-algorithm makes critical use of certain systematic sampling techniques from statistics (viz.\ line $4$ of Algorithm \ref{sage}).
 Consider the problem of sampling without replacement where one needs to randomly sample a $k$-set $S$ from the universe $[N]$ such that item $i \in [N]$ is included in the set $S$ with a pre-specified marginal inclusion probability $p_i,~ \forall i \in [N].$ Formally, if the $k$-set $S$ is sampled with probability $\mathbb{P}(S),$ we require that 
$ \sum_{S: i \in S, |S|=k} \mathbb{P}(S) = p_i, \forall i \in [N]. $
 Since the sampling is done without replacement, for any $k$-set $S$, we have:
$\sum_{i \in [N]} \mathds{1}(i \in S) = k.$	
 Taking expectation of both sides with respect to the randomness of the sampler, we conclude that 
 any feasible marginal inclusion probability vector $\bm{p}$ must belong to the set $\Delta^k_N$ defined as follows:
 \begin{eqnarray} \label{nec_cond}
 	\sum_{i \in [N]} p_i = k, ~~\textrm{and} ~~0\leq p_i\leq 1, \forall i \in [N].	
 \end{eqnarray}
 It turns out that condition \eqref{nec_cond} is also \emph{sufficient} for designing efficient sampling schemes that leads to the marginal inclusion probability vector $\bm{p}.$ Such sampling schemes have been extensively studied in the statistical sampling literature under the heading of \emph{unequal probability sampling design} \citep{tille1996some, hartley1966systematic, hanif1980sampling}. 
  In this paper, we use a linear-time exact sampling scheme proposed by \citet{madow1949theory} as outlined below in Algorithm \ref{uneq}. 

\begin{algorithm} 
\caption{Madow's Sampling Scheme}
\label{uneq}
\begin{algorithmic}[1]
\REQUIRE A universe $[N]$ of size $N$, cardinality of the sampled set $k$, and a  marginal inclusion probability vector $\bm{p} = (p_1, p_2, \ldots, p_N)$ satisfying condition \eqref{nec_cond}  
\ENSURE A random $k$-set $S$ with $|S|=k$ such that, $\mathbb{P}(i \in S)=p_{i}, \forall i\in [N]$ 

\STATE Define $\Pi_0=0$, and $\Pi_i= \Pi_{i-1}+p_{i}, \forall 1\leq i \leq N.$
\STATE Sample a uniformly distributed random variable $U$ from the interval $[0,1].$
\STATE $S \gets \emptyset$
\FOR {$i\gets 0$ to $k-1$}
\STATE Select the element $j$ if $\Pi_{j-1} \leq U + i < \Pi_j.$ 
\STATE $S \gets S \cup \{j\}.$
\ENDFOR
\RETURN $S$
\end{algorithmic}
\end{algorithm}
\paragraph*{Correctness:} The correctness of Madow's sampling scheme is easy to establish. From the necessary condition \eqref{nec_cond}, it follows that Algorithm \ref{uneq} selects exactly $k$ elements. Furthermore,  the element $j$ is selected if the random variable $U \in \sqcup_{i=1}^{N}[\Pi_{j-1} -i, \Pi_j-i ).$ Since $U$ is uniformly distributed in $[0,1]$, the probability that the element $j$ is selected  is equal to $\Pi_j - \Pi_{j-1} = p_j, \forall j\in [N].$ 

\section{FUNDAMENTAL LIMITS OF ONLINE PREDICTION WITH $k$-sets} \label{cover}
Consider the canonical binary prediction problem studied by \cite{cover1966behavior}. Assume that an adversary secretly selects a binary sequence $\bm{y}=(y_1, y_2, \ldots, y_T)$. The sequence is revealed to the learner one symbol at a time according to the following protocol - upon seeing the  initial segment of the sequence $y_1^{t-1}\equiv (y_1, y_2,\ldots, y_{t-1})$ at time $t$, the learner makes a (randomized) guess $\hat{y}_t$ for the $t$\textsuperscript{th} element of the sequence $y_t$. The actual value of $y_t$ is then revealed to the learner after the prediction. Let $\mu_{\mathcal{A}}(\bm{y})$ denote the fraction of mistakes made by a randomized prediction algorithm $\mathcal{A}$ for the sequence $\bm{y}$, \emph{i.e.,}
$\mu_{\mathcal{A}}(\bm{y})= \mathbb{E}^{\mathcal{A}}[T^{-1}\sum_{t=1}^T \mathds{1}(y_t \neq \hat{y}_t)],$
where the expectation is taken with respect to the randomness of the prediction algorithm. In Eqn.\ \eqref{cov-ext} below, we show that  irrespective of the prediction algorithm $\mathcal{A}$, the average fraction of errors $\mu_{\mathcal{A}}(\cdot)$ over all possible  $2^T$ binary sequences is precisely $\nicefrac{1}{2}.$ 
A loss function $\phi : \{\pm 1\}^T \to [0,1]$ is said to be \emph{achievable} if there exists an online prediction policy $\mathcal{A}$ such that the average prediction error under the policy $\mathcal{A}$ for any sequence is upper bounded by the function $\phi$, \emph{i.e.,} $\mu_{\mathcal{A}}(\bm{y}) \leq \phi(\bm{y}), \forall \bm{y}.$
An immediate question is to characterize the set of all achievable loss functions $\phi(\cdot)$. 

For a given sequence $\bm{y},$ let $\phi(\ldots, j, \ldots)$ be a shorthand for the quantity $\phi(y_1, y_2, \ldots, y_{t-1}, j, y_{t+1}, \ldots, y_T).$ 
We call a loss function  $\phi: \{\pm 1\}^T \to [0,1]$ to be \emph{stable} if it satisfies the following inequality for all  $\bm{y} \in \{\pm 1\}^T$ and for all time index $1\leq t \leq T$: 
\begin{eqnarray} \label{stable-loss}
 | \phi(\ldots, \underbrace{+1}_{t\textsuperscript{th} \textrm{ coordinate}}, \ldots)- \phi(\ldots,\underbrace{-1}_{t\textsuperscript{th} \textrm{ coordinate}}, \ldots) | \nonumber \\
 \leq \frac{1}{T}. 
 \end{eqnarray}
In this setup, \citet{cover1966behavior} proved the following result:
\begin{theorem}[Cover'66] \label{CoverTh}
	Suppose the loss function $\phi: \{\pm 1\}^T \to [0,1]$ is stable. Then $\phi (\cdot)$ is achievable if and only if $\mathbb{E}\phi (\bm{z})\geq  \nicefrac{1}{2},$ where the expectation is taken with respect to the i.i.d.\ uniform distribution over $\{\pm 1\}^T$.  
\end{theorem}
We emphasize that although the statement of Theorem \ref{CoverTh} involves an expectation, \emph{no} probabilistic assumption was made on the sequence $\bm{y}$. \citet{rakhlin2016tutorial} extended Theorem \ref{CoverTh} to the $N$-ary setting. 
In this paper, we generalize the result further to the \texttt{$k$-sets} setting where, instead of predicting a single value $\hat{y}_t,$ the learner is allowed to predict a (randomized) subset $S_t \subseteq [N]$ containing $k$ elements. 
Thus, the average loss incurred by a prediction policy $\mathcal{A}$ for the sequence $\bm{y}$ is given by:
\begin{eqnarray} \label{lossfn}
	\mu_{\mathcal{A}}(\bm{y})= \mathbb{E}^{\mathcal{A}}\left[\frac{1}{T}\sum_{t=1}^T \mathds{1}( y_t \notin S_t) \right],
\end{eqnarray}
where the expectation is taken with respect to the randomness of the policy $\mathcal{A}$. 
 Uniformly averaging the loss function $\mu_{\mathcal{A}}(\cdot)$ over all $N^T$ possible $N$-ary sequences $\bm{y}$ (equivalently, endowing the set of all sequences in $[N]^T$ the i.i.d. uniform probability measure), we have 
 \begin{align} \label{cov-ext}
 \mathbb{E} \mu_{\mathcal{A}}(\bm{y})&= \mathbb{E}\mathbb{E}^{\mathcal{A}}\big[\frac{1}{T}\sum_{t=1}^T \mathds{1}( y_t \notin S_t) \big] \nonumber \\
 &\stackrel{(\textrm{Fubini's Th.})}{=} \mathbb{E}^{\mathcal{A}}\mathbb{E}\big[\frac{1}{T}\sum_{t=1}^T \mathds{1}( y_t \notin S_t) \big] \nonumber \\
 &\stackrel{(a)}{=} 1-\frac{k}{N}, 
 \end{align}  
where (a) follows from the fact that $|S_t|=k, \forall t.$  As in condition \eqref{stable-loss}, we call a loss function $\phi: [N]^T \to [0,1]$ to be \emph{stable} if for all sequences $\bm{y} \in [N]^T$ and all coordinates $t$ of $\phi$ the following two conditions hold:
 \begin{align} 
&\max_{i\in [N]} \phi (\ldots, i, \ldots) - \frac{1}{N}\sum_{j \in [N]} \phi(\ldots, j, \ldots) \leq \frac{k}{NT}, \label{sta1}\\
& \frac{1}{N}\sum_{j \in [N]} \phi(\ldots, j, \ldots) - \min_{i\in [N]} \phi (\ldots, i, \ldots) \leq \big(1-\frac{k}{N}\big)\frac{1}{T}.\label{sta2}
\end{align}
Our first result generalizes Cover's theorem by showing that conditions \eqref{sta1} and \eqref{sta2} together are also sufficient for the achievability. 
 \begin{theorem} \label{main_thm}
Suppose the loss function $\phi: [N]^T \to [0,1]$ is stable. Then $\phi (\cdot)$ is achievable by some online policy if and only if $\mathbb{E}\phi (\bm{z})\geq 1-\nicefrac{k}{N},$ where the expectation is taken w.r.t. the i.i.d.\ uniform distribution over $[N]^T.$  	
 \end{theorem}
The necessity part of Theorem \ref{main_thm} has already been established in Eqn.\ \eqref{cov-ext} above. The proof of sufficiency is constructive and proceeds in two phases. In Phase-I, at each round $t$, we compute a vector $\bm{p}_t$ satisfying the feasibility condition \eqref{nec_cond}, such that $p_{ti}$ gives the correct marginal inclusion probability of the element $i \in [N]$ that achieves the loss function $\phi(\cdot)$. 
In Phase-II, we sample a $k$-set $S_t \subseteq [N]$ according to the marginal inclusion probabilities $\bm{p}_t$ using Algorithm \ref{uneq}. Please refer to Section  \ref{main_thm_proof} of the supplementary material for the proof of Theorem \ref{main_thm}. 

 \paragraph{Discussion:} It is to be noted that directly using the generic online policy appearing in the achievability proof of Theorem \ref{main_thm} could be intractable in terms of computation or memory requirements. A more serious issue with the generic prediction policy is that it requires the loss function to be \emph{stable}, which limits its applicability. 
 Similar to the treatment in \citet{rakhlin2016tutorial}, it might be possible to work with some relaxation of the loss function to derive a tractable policy. In the rest of the paper, we show that near-optimal inclusion probabilities may be efficiently computed via alternative methods, which result in low-regret efficient online prediction policies.

\section{LEARNING POLICIES FOR THE \texttt{$k$-sets} PROBLEM} \label{hedge-k-set}
In this section, we propose two different efficient online policies for the \texttt{$k$-sets} problem. The first policy uses \texttt{Hedge} as the base policy and the second policy utilizes the standard \emph{Follow-the-Regularized-Leader} framework. 
\subsection{\texttt{$k$-sets} with \texttt{Hedge}} \label{hedge-sec}
For the simplicity of exposition, we use the  the standard \texttt{Hedge} policy as our base policy in conjunction with the \texttt{SAGE} meta-algorithm. It will be clear from the sequel that any other \texttt{Experts} policy, such as \texttt{Squint} \citep{koolen2015second} or \texttt{AdaHedge} \citep{erven2011adaptive}, may also be used as the base policy, leading to more refined regret bounds. 

\paragraph{1. The Base Policy:}  
We start with the standard meta-experts framework as discussed in Section \ref{intro1}. Define a collection of $\binom{N}{k}$ experts, each corresponding to a distinct $k$-subset of the set $[N]$. Assume that the learner predicts the set $S$ with probability $p_t(S), \forall S \in \binom{[N]}{k}$. The expected reward accrued by the learner when the adversary chooses symbol $y_t$ at time $t$ is given by:
\begin{align}
   &  \mathbb{E}\big[\sum_{S: y_t\in S} 1\times \mathds{1}(S_t=S) + \sum_{S: y_t \notin S} 0 \times \mathds{1}(S_t=S)\big]\nonumber \\
   &= \mathbb{P}(y_t \in S_t) = p_t(y_t),
\end{align}
where $p_t(i) := \sum_{S: i \in S}p_t(S)$ is the marginal inclusion probability of the $i$\textsuperscript{th} element in the predicted $k$-set $S$. We now use the \texttt{Hedge} policy as our base policy for the resulting \texttt{Experts} problem. Let the indicator variables $r_\tau(i) := \mathds{1}(y_\tau=i), \forall i$ encode the symbol chosen by the adversary at round $\tau$. Furthermore, let the variable $r_{\tau}(S) := \sum_{i\in S}r_{\tau}(i)$ denote the reward accrued by the expert $S$ at round $\tau$. The cumulative reward accumulated by the expert $S$ up to the round $t-1$ is given by $R_{t-1}(S) = \sum_{\tau = 1}^{t-1} r_\tau(S).$ Overloading the notations a bit, let the variable $R_{t-1}(i)$ denote the number of times the $i$\textsuperscript{th} element appears in the sub-sequence $\bm{y}_1^{t-1}$. The \texttt{Hedge} policy with learning rate $\eta > 0$ chooses the expert $S$ at round $t$ with the following probability \citep{freund1997decision, vovk1998game}:
\begin{align}
    \label{eq:naive-Hedge-updates}
    p_{t}(S) & = \frac{w_{t-1}(S)}{\sum_{S'\subseteq [N]:\abs{S'}=k} w_{t-1}(S')}, ~~~ \forall S \in \binom{[N]}{k},
\end{align}
where $w_{\tau}(S) := \exp(\eta R_{\tau}(S))$. 
    
    \paragraph{2. Efficient Computation of the Inclusion Probabilities:}
 The marginal inclusion probabilities for each of the $N$ elements can be obtained by marginalizing the joint distribution given by Eqn.\ \eqref{eq:naive-Hedge-updates}. Let $w_{t-1}(i) := \exp(\eta R_{t-1}(i)).$ We have  
\begin{align}
    \label{eq:Hedge-marginal-file-selection-probabilities}
     p_{t}(i) 
    & = \sum_{S:\abs{S}=k,i\in S}p_t(S) \nonumber \\
    &= \frac{w_{t-1}(i)\sum_{S\subseteq [N]\setminus \{i\}: \abs{S} = k-1}w_{t-1}(S)}{\sum_{S'\subseteq [N]:\abs{S'}=k} w_{t-1}(S')},
\end{align}
where we have used the fact that for any $S \subseteq [N] \setminus \{i\}$, we have
   $w_{t-1}(i) w_{t-1}(S) = w_{t-1}(S \cup \{i\}).$
Clearly,
 \begin{align}
    \sum_{i\in [N]}p_t(i) & = \frac{\sum_{i\in [N]} w_{t-1}(i)\sum_{S\subset [N]\setminus \{i\}: \abs{S} = k-1}w_{t-1}(S)}{\sum_{S'\subset [N]:\abs{S'}=k} w_{t-1}(S')} \nonumber
     \\ &\stackrel{(a)}{=} k,
    \end{align}
    where step (a) follows from the fact that for any \texttt{$k$-set} $S$, the term $w_{t-1}(S)$ appears in the numerator exactly $k$ times. Therefore, the marginal inclusion probabilities in Eqn.\ \eqref{eq:Hedge-marginal-file-selection-probabilities} satisfy the feasibility condition \eqref{nec_cond}. Hence, given the marginal inclusion probabilities, Algorithm \ref{uneq} may be used to efficiently sample the predicted $k$-set. However, naively computing the marginal inclusion probabilities using Eqn.\ \eqref{eq:Hedge-marginal-file-selection-probabilities} requires evaluating sums of $\binom{N-1}{k-1}$ terms, which is  computationally intractable. This difficulty can be alleviated upon realizing that both the numerator and denominator of Eqn.\ \eqref{eq:Hedge-marginal-file-selection-probabilities} can be expressed in terms of elementary symmetric polynomials as shown below. For any vector $\bm{w} = (w_1, w_2, \ldots, w_N) \in \mathbb{R}^N,$ define the associated \emph{elementary symmetric polynomial} (ESP) of order $l$ as: 
         \begin{eqnarray} \label{elementary-symmetric}
         e_l(\bm{w}) = \sum_{I \subseteq [N], |I|=l} \prod_{j \in I} w_j.	
         \end{eqnarray}
 Furthermore, for any index $i \in [N], $ let $\bm{w}_{-i} \equiv (w_1, \ldots, w_{i-1}, w_{i+1}, \ldots, w_{N})\in \mathbb{R}^{N-1}$ denote the sub-vector with its $i$\textsuperscript{th} component removed. Then, from Eqn.\ \eqref{eq:Hedge-marginal-file-selection-probabilities}, it follows that $p_t(i)= \frac{w_{t-1}(i)e_{k-1}(\bm{w}_{{t-1}, -i})}{e_k(\bm{w}_{t-1})}.$ Hence, the marginal inclusion probabilities can be expressed in terms of symmetric polynomials that can be efficiently computed in $O(N \log^2(k))$ time via Fast Fourier Transform methods (see, \emph{e.g.,} \citet{shpilka2001depth}). Further speedup is possible by exploiting the fact that the weight of only one of the components change at a round. This faster iterative method is derived in Section \ref{sec:efficient-implementation-online-hedge} of the supplementary material. 
 \paragraph{3. Sampling the predicted set: } Upon computing the marginal inclusion probabilities, we use Madow's systematic sampling scheme outlined in Algorithm \ref{uneq} to sample a $k$-set. The overall prediction policy is summarized in Algorithm \ref{hedge-algo}. 
 \begin{algorithm}
  \caption{\texttt{$k$-sets} via \texttt{SAGE} with $\pi_{\textrm{base}}=$ \texttt{Hedge}}
  \label{hedge-algo}
 \begin{algorithmic}[1]
 \REQUIRE $\bm{w} \gets \bm{1},$ learning rate $\eta > 0.$
 \FOR {every time $t$}
 \STATE $\bm{w}_i \gets \bm{w}_i \exp(\eta \mathds{1}(y_{t-1}=i)), \forall i \in [N].$
 \STATE $p(i)\gets \frac{w(i)e_{k-1}(\bm{w}_{-i})}{e_k(\bm{w})}, \forall i \in [N],$ 
 \STATE Sample a $k$-set with the marginal inclusion probabilities $\bm{p}$ using Algorithm \ref{uneq}.
 \ENDFOR	
 \end{algorithmic}
 \end{algorithm}

 \subsubsection{Regret Bounds}
Recall that, in expectation, the performance of Algorithm \ref{hedge-algo} and the base policy \texttt{Hedge} are identical. It is well-known that by adaptively tuning the learning rate $\eta $, the \texttt{Hedge} policy with $n$ experts admits the following data-dependent small-loss regret bound \citep{koolen2010hedging, erven2011adaptive}
 \begin{eqnarray} \label{hedge_regret_bound}
 \textrm{Regret}_T \leq 	\sqrt{2l^*_T \ln n}+ \ln n,
 \end{eqnarray}
where $l^*_T$ denotes the cumulative loss incurred by the best fixed expert in hindsight for the given loss matrix. In the case of the \texttt{$k$-sets} problem, the total number of experts is given by $n = \binom{N}{k}\leq (\frac{Ne}{k})^k.$ Hence, the \texttt{SAGE} prediction framework with \texttt{Hedge} as the base policy yields the following adaptive regret bound:
\begin{eqnarray} \label{small-loss}
	\textrm{Regret}_T(\bm{y}) \leq \sqrt{2kl_T^*(\bm{y})\ln(Ne/k)} + k \ln(Ne/k),
\end{eqnarray}
 where $l_T^*(\bm{y})$ is the number of mistakes incurred by the best fixed $k$-set in hindsight for the sequence $\bm{y}$. Since $l^*_T(\bm{y}) \leq T$, the regret upper bound \eqref{small-loss} is sublinear in the horizon-length. However, the bound could be much smaller if the offline oracle incurs a small number of mistakes for a particular sequence. 
 
 \paragraph*{Discussion:} Algorithm~\ref{hedge-algo} offers a new projection and decomposition-free approach to break the existing $O(N^2)$ complexity barrier for the \texttt{$k$-sets} problem~\citep{herbster2001tracking}. The work by {\cite{uchiya2010algorithms}} studies a bandit version of the \texttt{$k$-sets} problem and proposes \textbf{Exp3.M} policy, which incurs $O(\sqrt{kNT \log N/k})$ regret. However, this bound cannot be compared with our (smaller) regret bound, applicable in the full-information setting. Furthermore, they use \emph{dependent rounding} method, which is more complex than Madow's sampling that we use here.
 \subsection{\texttt{$k$-sets} with \texttt{FTRL}} \label{ftrl}
 It is also possible to design efficient online policies for the \texttt{$k$-sets} problem with a base policy other than \texttt{Hedge}. 
 In Section \ref{oco-sec} of the supplementary, we show how the standard \emph{Follow-the-Regularlized-Leader} (\texttt{FTRL}) framework can be augmented with the systematic sampling schemes to design an efficient online prediction policy for a generalized version of the \texttt{$k$-experts} problem with the \texttt{sum-reward} function. 
 A drawback of the \texttt{FTRL} approach is that, unlike \texttt{Hedge}, this policy does not admit an adaptive regret bound. Due to space constraints, we defer the detailed discussion to Section \ref{oco-sec} of the supplementary material. 

\section{ $k$-\texttt{experts} WITH \texttt{Pairwise-rewards} } \label{pairwise_section}
In this section, we design an online prediction policy for a special case of the \texttt{$k$-experts} problem with the  \texttt{pairwise-reward} function and binary rewards (see Table \ref{k-experts-variants})\footnote{The general case with arbitrary rewards can be handled using a similar FTRL approach as in Section \ref{ftrl}.}. Recall that, in the \texttt{$k$-sets} problem, the adversary chooses a single item at each round (so that only one component of the reward vector $\bm{r}_t$ is one and the rest are zero). On the contrary, in this problem, the adversary secretly selects a \emph{pair} of items at each round (so that exactly two components of the reward vector $\bm{r}_t$ are one and the rest are zero). If \emph{both} the items chosen by the adversary are included in the predicted $k$-set, the learner receives a unit reward; else, it receives zero rewards for that round.  
The following hardness result is immediate.
\begin{proposition} \label{pair-NP-hard}
The offline version of the $k$-\texttt{experts} problem with \texttt{pairwise-rewards} is \textbf{NP-Hard}.	
\end{proposition}

\begin{proof}
The proof follows from a simple reduction of the \textbf{NP-Hard} \textsf{Densest $k$-subgraph} problem \citep{sotirov2020solving} to the offline optimization problem. 
Consider an arbitrary graph $\mathcal{G}$ on $N$ vertices and $T$ edges denoted by $e_1, e_2, \ldots, e_T$. Construct an instance of the \texttt{$k$-experts} problem with \texttt{pairwise-rewards} such that, at round $t$, the adversary chooses the pair of items corresponding to the vertices of the edge $e_t, 1\leq t \leq T.$
Then the problem of finding a subgraph of $k$ vertices such that the number of edges in the induced subgraph is maximum (\emph{i.e.,} the \textsf{Densest $k$-subgraph} of $\mathcal{G}$) reduces to the offline problem of selecting the most rewarding $k$ items to maximize the cumulative reward in the $k$-\texttt{experts} problem with \texttt{pairwise-rewards}.   
\end{proof} 

In principle, we can use the \texttt{SAGE} framework to obtain the optimal pairwise inclusion probabilities and then sample $k$ items accordingly. However, there are two main difficulties with this approach - (1) unlike Eqn.\ \eqref{nec_cond}, there is no known succinct characterization of the feasible set of pairwise inclusion probability vector when $k$ items are chosen from $N$ items without replacement, and (2) given a feasible pairwise inclusion probability vector, it is not known how to efficiently sample $k$ items accordingly. The above roadblocks are not surprising given the hardness of the offline problem. This prompts us to propose the following approximate policy described in Algorithm \ref{pairwise-algo}.

 \begin{algorithm}
  \caption{Algorithm for \texttt{pairwise-rewards}}
  \label{pairwise-algo}
 \begin{algorithmic}[1]
 \STATE Treat each pair of items as a single \emph{super-item}. 
 \STATE Use \textsf{SAGE} to sample $k$ distinct super-items from $\binom{N}{2}$ super-items  per round.
 \end{algorithmic}
 \end{algorithm}
%
Since any particular item may be a part of $k-1$ super-items, it is possible that the set of sampled super-items in Algorithm \ref{pairwise-algo} includes an item multiple times. However, it is easy to see that the number of items contained in the union of any $k$ super-items is bounded between $\sqrt{2k}$ and $2k$. Hence, replacing $N$ with $\binom{N}{2}$ (the number of super-items) in Eqn.\ \eqref{small-loss} yields the following performance guarantee for Algorithm \ref{pairwise-algo}:
 
Offline oracle reward with at most $\sqrt{2k}$ items - the reward accrued by Algorithm \ref{pairwise-algo} with at most 
 $2k$ items is upper bounded by:
\[ 2\sqrt{kl_T^*\ln(N^2e/2k)} + 2k \ln(N^2e/2k),\]
where $l_T^*$ is the loss incurred by the optimal offline oracle using $2k$ items.	
Algorithm \ref{pairwise-algo} is an instance of \emph{improper learning} algorithm where the online policy competes with a weaker oracle. 
\section{LEARNING POLICIES FOR MONOTONE REWARDS} \label{general_reward}
In this section, we use the \texttt{SAGE} framework to design an efficient online policy to learn any smooth monotone reward function. Recall that a set function $f: 2^{[N]} \to \R$ is \emph{monotone} if $f(S_1) \geq f(S_2), \forall S_2 \subseteq S_1 \subseteq [N].$ A set function $f$ is \emph{modular} if 
 for any subset $S \subseteq [N],$ we have:
$f(S) = \sum_{i\in S}f(\{i\}).$
Our starting point is the following fundamental result, which approximates \emph{any} set function by modular functions.
\begin{theorem}[\cite{iyer2012algorithms}]\label{thm:sandwich} For a given set $X$ and any set function $f: 2^X\to\R$ and any set $Y\subseteq X$, there are two modular functions $m_u: 2^X\to\R$ and $m_l: 2^X\to\R$ such that $m_l\leq f\leq m_u$ and $m_l(Y) = f(Y)=m_u(Y)$. Furthermore, the functions $m_l$ and $m_u$ can be expressed explicitly in terms of the function $f$. 
\end{theorem}
See Appendix \ref{monotone_reward} for the expressions of approximating modular functions and other computational details. We assume that the reward function $f_t$, chosen by the adversary at any round $t\in[T],$ is monotone with $f_t(\emptyset)=0,\forall t\in[T]$. We also assume that the reward functions are ``smooth'', i.e., there exists a finite constant $G$ such that $\forall S \subseteq [N], x \in [N]$, we have:
\begin{eqnarray} \label{smoothness}
|f_t(S)-f_t(S\setminus \{x\})|\leq G, ~ \forall t \geq 1. 
\end{eqnarray}
In the \texttt{$k$-experts} setting, the online prediction policy can select only a subset of $k$ experts at each round.  
We consider an improper learning setup where our objective is to design a prediction policy that attains at least a $\nicefrac{k}{N}$ fraction of the total cumulative rewards obtained by taking \emph{all} $N$ experts at each round up to an $O(\sqrt{T})$ term. Note that the comparator in this section is different from that of the standard regret metric \eqref{regret-def}, where the reward accrued by the online policy is compared against the optimal $k$-set in hindsight. 

Using Theorem \ref{thm:sandwich}, we can construct a modular set function $m_l^t$ corresponding to the function $f_t$ such that:
\begin{align} \label{main-cond}
    f_t\geq m_l^t, &&\text{and}&& f_t([N])=m_l^t([N]).
\end{align}
Consider a \texttt{sum-reward} variant of the \texttt{$k$-sets} problem, where the reward $g_t(i)$ for the $i$\textsuperscript{th} expert at round $t$ is set to be equal to $m_l^t(\{i\}), i\in [N].$
We now use a prediction policy that minimizes the static regret \eqref{regret-def} with respect to the linearized reward vectors $\{\bm{g}_t\}_{t\geq 1}$: 
\begin{align*}\label{eq:regret}
    \mathcal{R}_T = \max_{p^*\in \Delta_N^k}\sum_{t\leq T}\langle g_t, p^*\rangle -\sum_{t\leq T} \langle g_t, p_t\rangle.
\end{align*}
From Eqn.\ \eqref{reg-bd-main} of the Supplementary, it follows that the \texttt{FTRL} ($\eta$) policy with entropic regularizer guarantees the following regret bound for the \texttt{sum-reward} problem:
\begin{align*}
    \mathcal{R}_T \leq \frac{k\ln (N/k)}{\eta} + 2\eta\sum_{t\leq T}\norm{\bm{g}_t^2}_{k, \infty},
\end{align*}
where $\norm{\bm{x}^2}_{k, \infty}$ denotes the sum of the $k$ largest components of the vector $(x_1^2, x_2^2,\ldots, x_N^2)$.
Using the smoothness assumption \eqref{smoothness}, we show in Appendix \ref{monotone_reward} that $\norm{\bm{g}_t^2}_{k, \infty} \leq B^2$, where $B= O(GN^2\sqrt{k})$ for arbitrary reward functions. We also show that the bound can be improved to $B=O(G\sqrt{k})$ for submodular functions. Hence, with the optimal tuning of the learning rate $\eta,$ the \texttt{FTRL} policy achieves the following regret bound: 
\begin{align}
    \mathcal{R}_T \leq 2B\sqrt{2kT\ln(N/k)}.
\end{align}
Now observe that:
\begin{align*}
    \E[f_t(S_t)] &= \sum_{S_t} p_t(S_t)f_t(S_t)\geq \sum_{S_t}p_t(S_t)m_l^t(S_t)\\
    &=\sum_{i=1}^N p_t(i)g_t(i)=\langle g_t, p_t\rangle.\stepcounter{equation}\tag{\theequation}\label{eq:part1}
\end{align*}
 Furthermore, we also have: 
    \begin{align*}
    \sum_{t\leq T} f_t([N]) \stackrel{(a)}{=} \sum_{t\leq T}\sum_{i=1}^N g_t(i)
    \leq \frac{N}{k}\max_{p^*\in \Delta_N^k}\sum_{t\leq T}\langle g_t, p^*\rangle\stepcounter{equation}\tag{\theequation},\label{eq:part2}
\end{align*}
where we have used Eqn.\ \eqref{main-cond} in Eqn.\ $(a)$.
Substituting the bounds from Eqn.\ \eqref{eq:part1} and \eqref{eq:part2} into the regret bound \eqref{eq:regret} yields the following performance guarantee:
\begin{eqnarray*}
    \frac{k}{N}\sum_{t\leq T} f_t([N]) - \sum_{t\leq T}\E[f_t(S_t)]
    \leq2B\sqrt{2kT\ln(N/k)}. 
\end{eqnarray*}
 Hence, for arbitrary monotone reward functions, the prediction policy asymptotically achieves a $\nicefrac{k}{N}$ fraction of the maximum possible cumulative reward. 

\section{LOWER BOUNDS} \label{lower bound}
In this section, we lower bound the achievable regret for different variants of the $k$-\texttt{experts} problem.  
To begin with, consider the setting where the adversary chooses binary rewards with exactly one non-zero reward per round. 
 In this setting, \citet{sigmetrics20} established the following regret lower bound for the \texttt{Sum-reward} variant of the \texttt{$k$-experts} problem:
 \begin{theorem}[Regret Lower bound for \texttt{Sum-reward}] \label{sum-rew-lb}
 For any online policy with $\frac{N}{k} \geq 2$ and $T \geq 1$, we have
 	\[ \mathcal{R}_T^{\textrm{Sum-reward}} \geq \sqrt{\frac{kT}{2\pi}} -\Theta(\frac{1}{\sqrt{T}}).\]
 \end{theorem}
Note that with the above rewards structure, the \texttt{Sum-reward}, the \texttt{Max-reward}, and the \texttt{Pairwise-reward} variants of the $k$-\texttt{experts} problem become identical.  Hence, Theorem \ref{sum-rew-lb} also yields a lower bound to all of the above variants of the $k$-\texttt{experts} problem. However, from the standard \texttt{Hedge} achievability bound applied to the meta-experts (Eqn.\ \eqref{small-loss}), it can be readily observed that the upper and lower regret bounds differ by a logarithmic factor. Our main result in this section is the following tight regret lower bound for the \texttt{Max-Reward} variant of the \texttt{$k$-experts} problem, that removes the above logarithmic gap.
\begin{theorem}[Regret Lower Bound for \texttt{Max-reward}] \label{k-experts-lower-bound}
 For any online policy with $T \geq 16k \ln(\frac{N}{k})$ and $\frac{N}{k} \geq 7,$  we have
	\[ \mathcal{R}_T^{\texttt{Max-reward}} \geq 0.02\sqrt{kT\ln \frac{N}{k}}. \]
\end{theorem}
Compared to the standard lower bounds \citep{cesa2006prediction}, a distinguishing feature of the above regret lower bound is its non-asymptotic nature.   
%
 
\paragraph{Proof outline:} The proof utilizes the standard probabilistic technique where the worst-case regret is lower bounded by the average regret over an ensemble of \texttt{$k$-experts} problems. However, the analysis becomes complex as the reward accrued at each round $t$ is a non-linear function of the reward vector. 
To alleviate this difficulty, we first partition the pool of $N$ experts into $k$ disjoint subsets. Then we select the cumulative best expert in hindsight from each subset in order to lower bound the optimal offline reward. Please refer to Section \ref{k-experts-lower-bound-proof} in the supplementary material for detailed proof. 

\begin{figure*}[!htb]
\minipage{0.33\textwidth}
  \includegraphics[width=\linewidth]{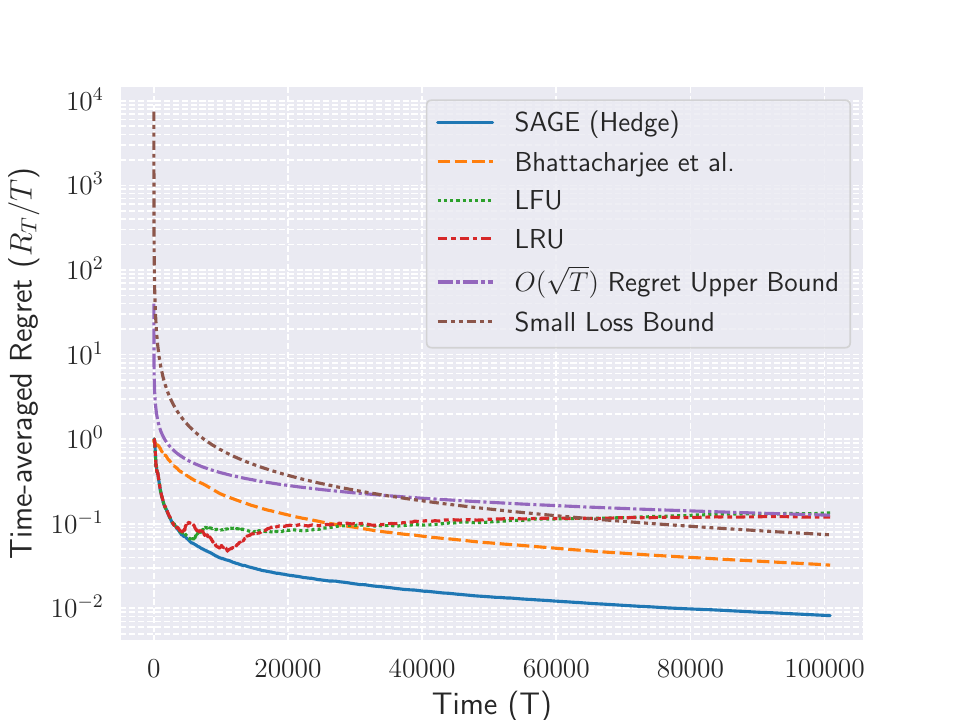}
  \caption{\small{Comparison among different \texttt{$k$-set} policies with $\nicefrac{k}{N}=0.1, N \sim 2400$ for the MovieLens Dataset.}}\label{fig:regret}
\endminipage\hfill
\minipage{0.33\textwidth}
  \includegraphics[width=\linewidth]{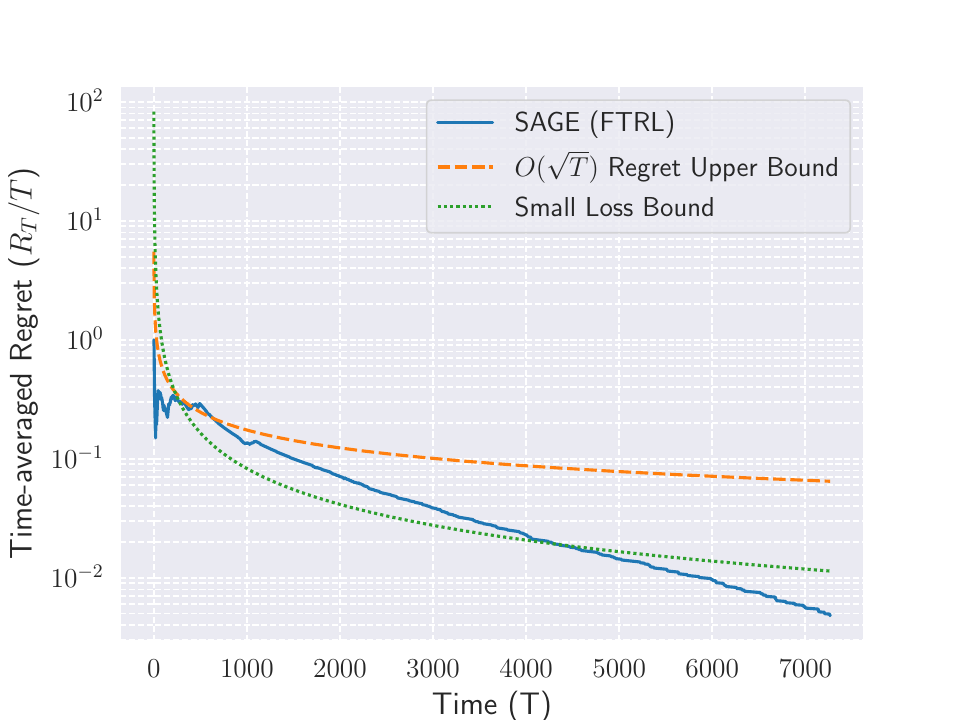}
  \caption{\small{Performance of the \texttt{SAGE} policy for pairwise predictions with $\nicefrac{k}{N}=0.02$ for the Reality Mining Dataset}.}\label{fig:regret_pairwise}
\endminipage\hfill
\minipage{0.33\textwidth}%
  \includegraphics[width=\linewidth]{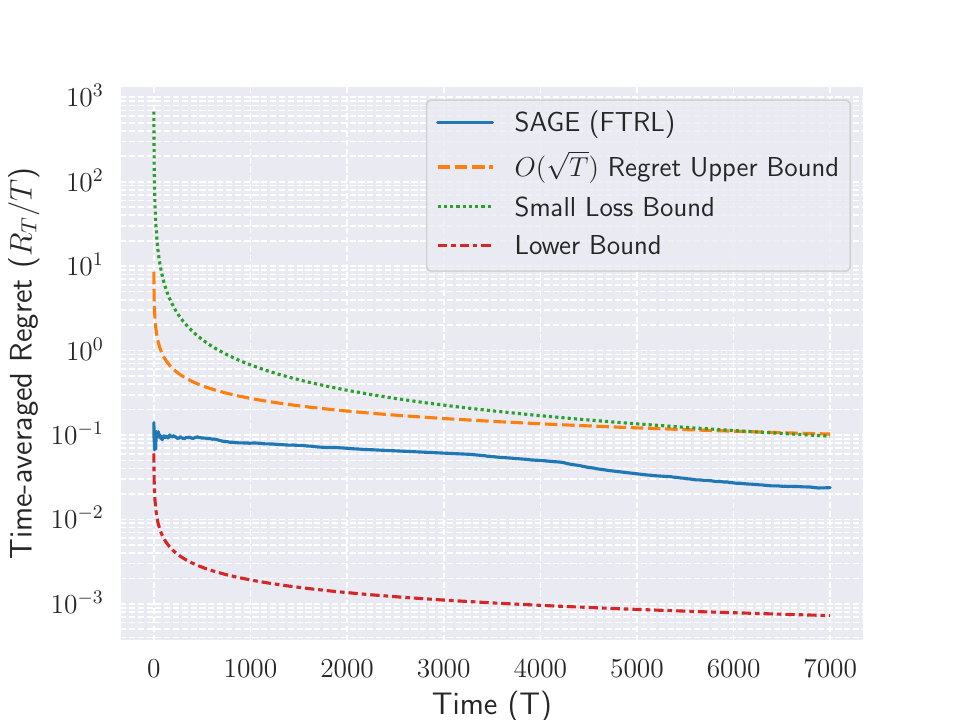}
  \caption{\small{Performance of \texttt{SAGE} for the $\texttt{Max-Reward}$ function, with $\nicefrac{k}{N}=0.01$ for the MovieLens Dataset.}}\label{fig:maxreward}
\endminipage
\end{figure*}

\section{NUMERICAL EXPERIMENTS} \label{sims}
\paragraph{\texttt{\textbf{$k$-sets:}}}\footnote{All codes used in the experiments are available at: \\\url{https://github.com/sourav22899/k-sets-problem}.} 
Assume that 
there is a collection of $N$ movies. The user may request any of the $N$ movies at each round. The learner sequentially predicts (possibly randomly) a set of $k$ movies that the user is likely to watch at a given round. At each round, the learner receives a unit reward if the movie chosen by the user is in the predicted set; else, it receives zero rewards for that round. The learner's goal is to maximize the total number of correct predictions over a given time interval. In our experiments, we use the MovieLens 1M dataset \citep{harper2015movielens} for generating the sequence of movies chosen by the user. The dataset contains $T\sim 10^5$ ratings for $N \sim 2400$ movies along with the timestamps. We assume that a user rates a movie immediately after watching it. The plot in Figure \ref{fig:regret} compares the normalized regrets of the proposed \texttt{SAGE} policy (with $\pi_{\textrm{base}}=$ \texttt{Hedge}), the \texttt{FTPL} policy proposed by \cite{sigmetrics20}, and two other baseline prediction policies -  LFU and LRU, which treat the prediction problem as a paging problem \citep{geulen2010regret}. 
From the plot, it is clear that the \texttt{SAGE} policy decisively outperforms all other policies.

 \paragraph{\textbf{\texttt{$k$-experts} with \texttt{Pairwise-reward}:}}  In our next experiment, we use the MIT Reality Mining dataset ~\citep{konect:eagle06} to understand the efficacy of the prediction policy for pairwise rewards proposed in Section \ref{pairwise_section}. The dataset contains timestamped human contact data among $100$ MIT students collected using standard Bluetooth-enabled mobile phones over $9$ months. In our experiments, we consider a subset of $N=20$ students with $\binom{20}{2}=190$ potential contact pairs. The learner's task is to predict a sequence of $k$-sets that include both the students involved in the contact for each timestamp. As described in Section \ref{pairwise_section}, we design an approximate prediction policy by considering each pair of students as a \emph{super-item} and use the $\texttt{SAGE}$ framework with $\pi_\text{base}=\texttt{FTRL}$. The normalized regret achieved by this policy is shown in Figure \ref{fig:regret_pairwise}. To compute the optimal static offline reward, we used a brute-force search. From the plots, we see that the normalized regret of this policy approaches zero for long-enough time-horizon. 
%
 \paragraph{\textbf{\texttt{$k$-experts} with \texttt{Max-reward}:}}
In our final experiment, we use a subset of the MovieLens dataset with  $T\sim 7000$ ratings for $N=200$ movies. We assume that the movies are sorted according to genres so that if the movie $i$ is chosen by the user at each round, the learner receives a reward of $\max_{j\in S}\left(1-\frac{1}{N}|j-i|\right)$ for predicting the set $S$. This reward function roughly emulates the practical requirement that if the requested movie is not in the predicted set, then it is preferable to recommend a similar movie than a completely different one. In Figure~\ref{fig:maxreward}, we plot the normalized regret of the $\texttt{SAGE}$ policy with $\pi_\text{base}=\texttt{FTRL},$ along with the lower bound given in Theorem \ref{k-experts-lower-bound}. From the plot, we can see that the normalized regret shows a downward trend with $T$ even with the \texttt{FTRL} policy, albeit there is a non-trivial gap with the lower bound. This gap is expected as the \texttt{FTRL} policy is optimal for the \texttt{Sum-reward} function, but not necessarily so for the \texttt{Max-reward} function. Please see Section \ref{addl-expts} of the supplement for additional results.
%



\section{CONCLUSION} \label{conclusion}
In this paper, we formulated the \texttt{$k$-experts} problem and designed efficient learning policies for some of its variants using the \texttt{SAGE} framework. We also derived a tight regret lower bound for the \texttt{Max-reward} variant and characterized the set of all mistake bounds for the \texttt{$k$-sets} problem achievable by online policies. In the future, it would be interesting to benchmark the performance of the algorithms on larger datasets.
\clearpage
\section{ACKNOWLEDGMENTS} \label{
ack}
This work is partially supported by the grant IND-417880 from
Qualcomm (USA) and a research grant from the Govt.\ of India under the Institutes of Eminence (IoE) initiative. The computational results
reported in this work were performed on the AQUA Cluster at
the High Performance Computing Environment of IIT Madras.
\balance
\bibliographystyle{unsrtnat}
\bibliography{bibmobility}
\clearpage
\onecolumn \makesupplementtitle
\section{Proofs and Derivations} \label{appendix}
\subsection{Proof of Theorem \ref{main_thm}} \label{main_thm_proof}
\paragraph*{Phase-I: Computation of the Marginal Inclusion Probabilities $\bm{p}_t:$}	
Similar to the treatment in \cite{rakhlin2016tutorial}, we use a potential function-based argument to derive a set of marginal inclusion probabilities at each time $t$ that leads to the loss function $\phi(\cdot)$. Let $\{\phi_t : [N]^t \to [0,1]\}_{t=0}^T$ be a sequence of potential functions satisfying the boundary condition 
\begin{eqnarray} \label{bd-cond}
\phi_T(\bm{y})= \phi(\bm{y}). 
\end{eqnarray}
We define $\phi_0$ to be a suitable constant. In order to achieve the loss function $\bm{\phi}(\cdot)$, we require the following equality to be valid for all sequences $\bm{y} \in [N]^T$:
\begin{eqnarray} \label{pred1}
	\mathbb{E}\bigg(\frac{1}{T}\sum_{t=1}^T \mathds{1}(y_t \notin S_t) \bigg) 
	 =\sum_{t=1}^T \big(\phi_t(\bm{y}^t) - \phi_{t-1}(\bm{y}^{t-1}) \big) + \phi_0, 
\end{eqnarray}
where the above equation follows from telescoping the summation and using the boundary condition \eqref{bd-cond}. 
For a given initial segment of the sequence $\bm{y}^{t-1},$ consider an online policy that includes the $i$\textsuperscript{th} element in the predicted set $S_t$ with the conditional probability $p_{ti}(\bm{y}^{t-1})$. Clearly  
\begin{eqnarray} \label{pred2}
\mathbb{P}(y_t \notin S_t| \bm{y}^{t-1}) = 1- \sum_{i=1}^N p_{ti}(\bm{y}^{t-1}) \mathds{1}(y_t=i).
\end{eqnarray}
Hence, combining equations \eqref{pred1} and \eqref{pred2}, the achievability is ensured if we can exhibit a sequence of potential functions $\{\phi_t(\cdot)\}$ and a randomized online strategy for selecting the sets $S_t$, such that the following equality holds for every sequence $\bm{y} \in [N]^T:$
\begin{eqnarray} \label{main_eq}
	 \sum_{t=1}^T\bigg( -\sum_{i=1}^N\frac{p_{ti}(\bm{y}^{t-1})\mathds{1}(y_t=i)}{T} + \phi_{t-1}(\bm{y}^{t-1}) 
	 - \phi_t(\bm{y}^t) +\frac{1}{T}(1-\phi_0) \bigg) = 0.
\end{eqnarray}
We now consider the following candidate sequence of potential functions:
\begin{eqnarray} \label{potential}
\phi_t(\bm{y}_t) \equiv \mathbb{E}\phi(\bm{y}_t, \epsilon_{t+1}^T),~~ \forall t,
\end{eqnarray}
 where the expectation is taken over a random sequence $\bm{\epsilon}_{t+1}^T$ such that each component $\epsilon_j, t+1\leq j \leq N$ is distributed i.i.d. uniformly over the set $[N].$ It is easy to see that, the boundary condition \eqref{bd-cond} is satisfied. Furthermore, from the condition given in the statement of the theorem, we have $\phi_0 = \mathbb{E} \phi (\bm{\epsilon}_{1}^T) = 1 - \nicefrac{k}{N}.$ Next, we exhibit a prediction strategy with inclusion probabilities $\{p_{ti}(\bm{y}^{t-1})\}$ such that the equation \eqref{main_eq} is satisfied. For, this, we set each of the terms of the equation \eqref{main_eq} identically to zero for any sequence $\bm{y} \in [N]^T.$ This yields the following conditional inclusion probability of the $i$\textsuperscript{th} element for any initial segment of the request sequence $ \bm{y}^{t-1} \in [N]^{t-1}:$
 \begin{align} \label{prob}
p_{ti}(\bm{y}^{t-1})  = T\bigg(\phi_{t-1}(\bm{y}^{t-1})-\phi_t(\bm{y}^{t-1}i)\bigg)  + \frac{k}{N}, ~~~ \forall i \in [N].	
 \end{align}
From the definition \eqref{potential}, we have that $\frac{1}{N}\sum_{i=1}^N \phi_t(\bm{y}^{t-1}i)= \phi_{t-1}(\bm{y}^{t-1}).$ Hence, summing equation \eqref{prob} over all $i \in [N],$ we have 
\begin{eqnarray*}
\sum_{i=1}^N p_{ti}(\bm{y}^{t-1}) = k.	
\end{eqnarray*}
Thus, the scalars $\{\bm{p}_{ti}\}_{i=1}^N$ satisfy the requirement in equation \eqref{nec_cond}. Hence, to guarantee that Eqn.\ \eqref{prob} yields a valid prediction strategy, we only need to ensure that $0\leq p_{ti} \leq 1, \forall i \in [N]$. In the following, we show that this requirement is also satisfied, thanks to the stability property of the loss function $\phi(\cdot).$ For this, we are required to ensure the following bound for all $\bm{y}^{t-1}:$
\begin{eqnarray} \label{bound}
	-\frac{k}{N} \leq T\bigg(\frac{1}{N}\sum_{i=1}^N \phi_t(\bm{y}^{t-1}i)-\phi_t(\bm{y}^{t-1}i)\bigg) \leq 1 - \frac{k}{N}.
\end{eqnarray}
It immediately follows that the stability conditions, given by equations \eqref{sta1} and \eqref{sta2}, are sufficient to ensure the bound in Eqn.\ \eqref{bound}.
\paragraph*{Phase-II: Sampling the Predicted set }  
We use the conditional marginal inclusion probabilities $\bm{p}_t$, derived in Eqn.\ \eqref{prob}, to construct a consistent randomized output set $S_t$ with $|S_t|=k$. Since the inclusion probabilities satisfy the feasibility constraints, we can use the Algorithm \ref{uneq} to construct the predicted set. 
Phase-I and Phase-II, taken together, complete the proof of the theorem.

\subsection{Iterative evaluation of the marginal inclusion probabilities}
    \label{sec:efficient-implementation-online-hedge}
    At any time $t$, consider the formal power series $g_t(X)$ defined as \begin{align}
        \label{eq:gt(X)-define}
        g_t(X)=\prod_{i\in [N]}(X - w_{t}(i)) = \sum_{j=0}^N a_{tj}X^j,
    \end{align}  i.e., $\forall j=0,\cdots, N,\ a_{tj}$ is the coefficient of $X^j$ in the expansion of $g_t(X)$. Then, by Vieta's formulae, we obtain, 
    \begin{align}
        \label{eq:at-define}
        \sum_{1\le i_1<i_2<\cdots<i_k\le N}\prod_{j=1}^k w_{t}(i_j) & = (-1)^{k} a_{t,N-k}\nonumber \\\Longleftrightarrow \sum_{S'\subset [N]:\abs{S'}=k} w_{t}(S') & = (-1)^{k} a_{t,N-k}.
    \end{align}
    Now define
    \begin{align}
    \label{eq:gti(X)-define}
        g_{ti}(X) = \frac{g_t(X)}{X - w_{t}(i)}=\sum_{j=0}^{N-1}b^{(i)}_{tj}X^j, 
    \end{align} 
    where $b_{tj}^{(i)}$ is the coefficient of $X^{j}$ in the expansion of $g_{ti}(X)$.
    Again using Vieta's formula, it follows that, 
    \begin{align}
    \label{eq:bt-define}
        \sum_{S\subset [N]\setminus \{i\}: \abs{S} = k-1}w_{t}(S) & = (-1)^{k-1}b^{(i)}_{t,N-k}.
    \end{align}
    Therefore, it follows that the probability selection rule~\eqref{eq:Hedge-marginal-file-selection-probabilities} can be expressed as below: 
    \begin{align}
    \label{eq:modified-marginals-step1}
        p_t(i) & = -\frac{w_{t-1}(i)b^{(i)}_{t-1,N-k}}{a_{t-1,N-k}},\ \forall i\in [N].
    \end{align}
    
    It now remains to find a computationally efficient way of updating the coefficients $a_{tj},b^{(i)}_{tj}$. To this direction, given the coefficients $\{a_{tj}\}_{j=0}^N$, we compute the coefficients $\{b_{tj}\}_{j=0}^{N-1}$ in the following way. Using the formal power series expansion $(1-X)^{-1} = \sum_{l\ge 0}X^l$, one can write, \begin{align}
        \label{eq:gti(X)-from-gt(X)}
        g_{ti}(X) & = -w_{t}^{-1}(i)g_{t}(X)\sum_{l\ge 0}X^lw^{-l}_{t}(i)\nonumber\\
        \ & = -w_{t}^{-1}(i) \sum_{j=0}^N \sum_{l=0}^{\infty} a_{tj}w_{t}^{-l}(i)X^{j+l}.
    \end{align}
    Therefore, $\forall 0\le j\le N-1$, \begin{align}
    \label{eq:btj-update-rule}
        b_{tj}^{(i)} & = -\sum_{l=0}^{j}a_{tl}w_{t}^{-(j-l+1)}(i)
    \end{align}
    Consequently, we can further express the probability selection rule from Eq.~\eqref{eq:modified-marginals-step1} as 
    \begin{align}
        \label{eq:pt-update}
        p_{t}(i) & = \frac{\sum_{j=0}^{N-k}a_{t-1,j}w_{t-1}^{-(N-k-j)}(i)}{a_{t-1,N-k}}. 
    \end{align}
We now proceed to find update rule for the coefficients $a_{tj}$. Let $f_t$ be the file requested at time $t$. Then, $R_{t}(f_t)=R_{t-1}(f_t) + \rho_t$, where $\rho_t = \mathds{1}(f_t\in S_t)$, whereas, $R_{t}(i)=R_{t-1}(i)$ if $i\ne f_t$. Therefore, \begin{align}
    \ & g_t(X) = \prod_{i=1}^N (X-w_{t}(i)) =g_{t-1}(X)\cdot \frac{X - w_{t}(f_t)}{X-w_{t-1}(f_t)}\nonumber\\
    \ & =g_{t-1,f_t}(X)(X-w_t(f_t)) =\sum_{j=0}^{N-1}b_{t-1,j}^{(f_t)}X^j(X-w_{t}(f_t)).
\end{align}
Therefore, using the above and the update rule of $b_{tj}^{(i)}$ from Eq.~\eqref{eq:btj-update-rule}, we obtain,
\begin{align}
    a_{tj} & = b^{(f_t)}_{t-1,j-1} - w_t(f_t) b_{t-1,j}^{(f_t)}\nonumber\\
    \ & = w_{t}(f_t)\sum_{k=0}^j a_{t-1,k}w_{t-1}^{-(j-k+1)}(f_t) 
    - \sum_{k=0}^{j-1} a_{t-1,k}w_{t-1}^{-(j-k)}(f_t)\nonumber\\
        \label{eq:at-update-prelim}
    \ & = (e^{\eta}-1)\sum_{k=0}^{j}a_{t-1,k}w_{t-1}^{-(j-k)}(f_t) + a_{t-1,j},
\end{align}
where in the last step we have used the fact that $w_{t}(f_t)w_{t-1}^{-1}(f_t)=e^{\eta}$, since $f_t$ is the requested file at time $t$ and hence $R_t(f_t)=R_{t-1}(f_t) + 1$. The update Eq.~\eqref{eq:at-update-prelim} can be used to obtain a further simplified recurrence to the update of the coefficients $a_{t,i}$ as below:
\begin{align}
     a_{t,j}  
    & = (e^{\eta}-1)w_{t-1}^{-1}(f_t)\sum_{k=0}^{j-1}a_{t-1,k}w_{t-1}^{-(j-1-k)}(f_t) + e^{\eta} a_{t-1,j},\nonumber\\
    \label{eq:at-update-j-ge-1}
    \ & = w_{t-1}^{-1}(f_t)\big(a_{t,j-1} - a_{t-1,j-1}\big) + e^{\eta}a_{t-1,j},\ \forall 1\le j\le N,\\
    \label{eq:at-update-j-=0}
     a_{t,0}  & = e^{\eta}a_{t-1,0}.
\end{align}
Using the update equations of $\{a_{tj}\}_{j=1}^N$ and $\{p_t(j)\}$ from Eqs.~\eqref{eq:at-update-j-ge-1},~\eqref{eq:at-update-j-=0} and~\eqref{eq:pt-update} respectively, we have the following iterative numerical procedure for computing the marginal inclusion probabilities:
\begin{algorithm} 
\caption{Iterative Computation of the Marginal Inclusion Probabilities}
\label{algo:proposed-policy}
\begin{algorithmic}[1]
\REQUIRE Learning rate $\eta>0$, 
\INIT $\bm{R}_0=\bm{0},a_{0,j} = (-1)^{N-j}\binom{N}{j},\ \forall 0\le j\le N$.
\FOR {$t=1,\cdots, T$}
\STATE Compute $w_{t-1}(i)=\exp(\eta R_{t-1}(i))\ \forall i\in [N]$, and set 
$p_{t}(i) = \frac{\sum_{j=0}^{N-k}a_{t-1,N-k-j}w_{t-1}^{-j}(i)}{a_{t-1,N-k}},\ \forall i\in [N]$.
\STATE Sample a set $S_t\subset [N]$ with $\abs{S_t}=k$ according to Madow's systematic sampling using the probabilities $\{p_{t}(i)\}_{i\in [N]}$ and construct the vector $\bm{y}_t\in \{0,1\}^N$, such that $y_{t,i}=1(i\in S_t)$.
\STATE Observe the requested file index $f_t$ and update 
\begin{align*}
R_t(i)\gets R_{t-1}(i) + \mathds{1}(f_t=i).
\end{align*}
\STATE Update 
\begin{align*}
    a_{t,0} & \gets e^{\eta}a_{t-1,0}\\
    a_{t,j} & \gets w_{t-1}^{-1}(f_t)\big(a_{t,j-1} - a_{t-1,j-1}\big) 
     + e^{\eta}a_{t-1,j},\ 1\le j\le N.
\end{align*} 
\ENDFOR
\end{algorithmic}
\end{algorithm}


  
  \subsection{Generalized \texttt{$k$-sets} with \texttt{FTRL}} \label{oco-sec}
In this section, we design an efficient online policy for a generalized version of the \texttt{$k$-sets} problem where the reward per round is modulated using a non-decreasing concave function
$\psi : \mathbb{R}_{\geq \bm{0}} \to \mathbb{R}$, called the \emph{link function}. In particular, the reward of the learner at round $t$ is defined to be $\psi(\bm{r}_t \cdot \bm{p}_t)$. In the special case when $\psi(\cdot)$ is the identity function, we recover the standard \texttt{$k$-sets} problem. The notion of link functions is common in the literature on Generalized Linear Models \citep{filippi2010parametric, li2017provably}. Note that, although the reward function could be non-linear, it still depends only on the marginal inclusion probabilities of the elements, and hence the \texttt{SAGE} framework applies. 
Formally, the objective of the learner is to design an efficient online learning policy to minimize the \emph{static regret} with respect to an offline oracle (the best fixed $k$-set in the hindsight), \emph{i.e.,}
\begin{eqnarray} \label{regret_def2}
\mathcal{R}_T := \max_{\bm{p}^* \in \Delta(\mathcal{C}^N_k)} \sum_{t=1}^T\psi(\bm{r}_t \cdot \bm{p}^*)- \sum_{t=1}^T \psi(\bm{r}_t \cdot \bm{p}_t), 
\end{eqnarray}
%
We augment the well-known \emph{Follow-the-Regularized-Leader} (\textbf{FTRL}) framework with the Systematic Sampling scheme in Algorithm \ref{uneq} to design an efficient online policy for the generalized \textbf{$\bm{k}$-sets} problem with a sublinear regret.
Interestingly, we will see that, when specialized to the $\bm{k}$-\textbf{sets} problem, the \textbf{FTRL}-based approach yields a different policy from the \texttt{Hedge}-based Algorithm \ref{hedge-algo}. The problem of finding the optimal marginal inclusion probabilities to minimize the regret in Eqn.\ \eqref{regret_def2} is an instance of the Online Convex Optimization (OCO) problem \citep{hazan2019introduction}. We use the standard \emph{Follow-the-Regularized-Leader} (\textbf{FTRL}) paradigm to design an online prediction policy with sublinear regret. We refer the reader to \cite{hazan2019introduction} for an excellent introduction to the \textbf{OCO} framework in general, and the \textbf{FTRL} policy in particular. 

Recall that, in the general \textbf{FTRL} paradigm, the learner's action at time $t$ is obtained by maximizing the sum of the cumulative rewards (or a linear lower bound to it) upto time $t-1$ and a strongly concave regularizer $g: \Omega \to \mathbb{R},$ where $\Omega$ is the set of all feasible actions of the learner. For the \textbf{Generalized} $\bm{k}$-\textbf{sets} problem, the vector of marginal inclusion probabilities is constrained to be in the set $\Omega = \Delta^N_k$, where  
$\Delta^N_k = \{\bm{p} \in [0,1]^N: \sum_{i=1}^N p_i = k. \}$	
In the following, we choose the usual (Shannon) entropic regularizer as our regularization function, \emph{i.e.}, we take $g(\bm{p}) = -\sum_{i=1}^N p_i \ln p_i.$ This choice is motivated by the well-known fact that the entropic regularization yields the \texttt{Hedge} policy for the \texttt{Experts} problem (where $k=1$) \citep{hazan2019introduction}. In our numerical experiments, we also investigate the performance of the R\'enyi and Tsallis entropic regularizers of various orders \citep{amigo2018brief}.
Choosing the entropic regularizer leads to the following convex program for determining the marginal inclusion probabilities $\bm{p}_t$ at the $t$\textsuperscript{th} round:
\begin{eqnarray} \label{ftrl-opt}
\bm{p}_t = \arg\max_{\bm{p} \in \Delta^N_k} \bigg[\big(\sum_{s=1}^{t-1} \nabla_s)^T \bm{p} - \frac{1}{\eta}\sum_{i=1}^N p_i \ln p_i,\bigg]
\end{eqnarray}
where $\nabla_{s,i} \equiv r_{s,i} \psi'(\bm{r}_s^T \bm{p}_s)$ denotes the $i$\textsuperscript{th} component of the gradient vector. Using convex duality, the optimal solution to \eqref{ftrl-opt} may be quickly determined in $\tilde{O}(N)$ time as shown in Algorithm \ref{general-k-sets} below.

\begin{algorithm}
\caption{\texttt{FTRL} for the generalized $k$-\textbf{sets} problem with the entropic regularizer}
\label{general-k-sets}
\begin{algorithmic}[1]
\REQUIRE $\bm{R} \gets \bm{0},$ learning rate $\eta >0$
\FOR {every time step $t$:}
\STATE $\bm{R} \gets \bm{R} + \nabla_{t-1}.$
\STATE Sort the components of the vector $\bm{R}$ in non-increasing order. Let $R_{(j)}$ denote the $j$\textsuperscript{th} component of the sorted vector $j \in [N].$
\STATE Find the largest index $i^* \in [N]$ such that 
$ (k-i^*) \exp(\eta R_{(i^*)}) \geq \sum_{j=i^*+1}^N \exp(\eta R_{(j)}).$ 
\STATE Compute the marginal inclusion probabilities as $p_i = \min (1, K \exp(\eta R_{i}))$, where $K \equiv \frac{k-i^*}{\sum_{j=i^*+1}^N \exp(\eta R_{(j)})}.$
\STATE Using Algorithm \ref{uneq}, sample a $\bm{k}$-set with the marginal inclusion probabilities $\bm{p}.$	
\ENDFOR
\end{algorithmic}	
\end{algorithm}
See Section \ref{algo-derivation} below for the derivation of the Algorithm \ref{general-k-sets}. 
Interestingly, although for $k=1$, the Algorithm \ref{general-k-sets} is identical to \ref{hedge-algo}, for $k>1$, the algorithms are quite different. The regret guarantee for the \textbf{FTRL} policy \eqref{ftrl-opt} for the \textbf{Generalized} $\bm{k}$-\textbf{sets} problem follows immediately from the standard results on the regret bound for the \textbf{FTRL} policy for general \textbf{OCO} problems. The simplified regret bound is given in the following theorem. 
\begin{theorem}[Regret Bound] \label{ftrl-regret-bound}
	With the learning rate $\eta > 0$, the \textbf{FTRL} policy for the generalized $\bm{k}$-\textbf{sets} problem with the entropic regularizer ensures that 
	\begin{eqnarray*}
	\textrm{\emph{Regret}}_T \leq \frac{k\ln{N/k}}{\eta} + 2\eta \sum_{t=1}^T  ||\nabla_{t}^2||_{k,\infty}, 
	\end{eqnarray*}
where $||\nabla_{t,i}^2||_{k,\infty}$ denotes the sum of the $k$ largest components of the vector $\nabla_t^2,$ which is obtained by squaring the vector $\nabla_t$ component wise. 
\end{theorem}

\begin{proof}
 Recall the following general regret bound for the \textbf{FTRL} policy from Theorem 5.2 of \cite{hazan2019introduction}. For a bounded, convex and closed set $\Omega$ and a strongly convex regularization function $g: \Omega \to \mathbb{R},$ consider the standard \textbf{FTRL} updates, \emph{i.e.},
 \begin{eqnarray}\label{ftrl-updates}
  \bm{x}_{t+1} = \arg\max_{\bm{x} \in \Omega}\bigg[ \big(\sum_{s=1}^t \nabla_s^T\big) \bm{x} - \frac{1}{\eta}g(x), \bigg]   
 \end{eqnarray} 
 where $\nabla_s= \nabla f_t(\bm{x}_s), \forall s$. Then, as shown in \cite{hazan2019introduction}, the regret of the \textbf{FTRL} policy can be bounded as follows:
  \begin{eqnarray}\label{reg-bd-main}
  \textrm{Regret}^{\textrm{FTRL}}_T \leq 2 \eta \sum_{t=1}^T || \nabla_t||_{*,t}^2 + \frac{g(\bm{u})-g(\bm{x}_1)}{\eta},
  \end{eqnarray}
 where the quantity $|| \nabla_t||_{*,t}^2$ denotes the square of the dual norm of of the vector induced by the Hessian of the regularizer evaluated at some point $\bm{x}_{t+\nicefrac{1}{2}}$ lying in the line segment connecting the points $\bm{x}_t$ and $\bm{x}_{t+1}$. In the \textbf{Generalized $k$-set} problem, the Hessian of the entropic regularizer is given by the following diagonal matrix 
 \begin{eqnarray*}
 \nabla^2 g(\bm{p}_{t+\nicefrac{1}{2}}) = \textrm{diag}([p_1^{-1}, p_2^{-1}, \ldots, p_N^{-1}]).
 \end{eqnarray*}
 For a vector $v$, let $||v||_{k, \infty}$ denote the sum of its $k$ largest components. 
 With this notation, we can write \[|| \nabla_t||_{*,t}^2 = \sum_{i=1}^N p_i \nabla_{t,i}^2 \leq ||\nabla_t^2||_{k, \infty} , \] where we have used the fact that $0\leq p_i \leq 1$ and $\sum_i p_i=k.$ In the above, the vector $\nabla_t^2$ is obtained by squaring each of the components of the vector $\nabla_t.$
 
 To bound the second term in \eqref{reg-bd-main}, define a probability distribution $\tilde{\bm{p}} = \bm{p}/k.$ We have 
 \begin{eqnarray*}
 0 \geq g(\bm{p})  = \sum_i p_i \log p_i = - k \sum_i \tilde{p}_i \log \frac{1}{p_i}\\
  \stackrel{\textrm{(Jensen's inequality)}}{\geq} -k \log \sum_i \frac{\tilde{p}_i}{p_i} = -k \log \frac{N}{k}.   	
 \end{eqnarray*}
Hence, the regret bound in \eqref{reg-bd-main} can be simplified as follows:
\begin{eqnarray*}
	\textrm{Regret}^{\textrm{k-set}}_T \leq \frac{k}{\eta} \log \frac{N}{k}+ 2\eta \sum_{t=1}^T || \nabla_t^2||_{k,\infty}. 
\end{eqnarray*}
\end{proof}

\subsubsection{Derivation of Algorithm \ref{general-k-sets}} \label{algo-derivation}
Recall that, via Pinsker's inequality \citep{pinsker}, the entropic regularizer is strongly concave with respect to the $\ell_1$ norm. 
Thus, strong duality holds and the optimal solution to the problem \eqref{ftrl-opt} can be obtained by using the KKT conditions \citep{boyd}. To simplify the notations, denote the cumulative sum of the gradient vectors $\sum_{s=1}^{t-1}\nabla_s$ by the vector $\bm{R}_{t-1}.$  Thus, the problem \eqref{ftrl-opt} may be explicitly rewritten as follows: 

\begin{eqnarray*}
\max 	\sum_{i=1}^N p_i R_{t-1, i} - \frac{1}{\eta}\sum_{i=1}^N p_i \ln p_i
\end{eqnarray*}
subject to, 
\begin{eqnarray}
&&\sum_{i=1}^N p_i = k \label{cap}\\
&& p_i \leq 1, ~~\forall i \label{feas}\\
&& p_i \geq 0, ~~\forall i.	
\end{eqnarray}
By associating the real variable $\lambda$ with the constraint \eqref{cap} and the non-negative dual variable $\nu_i$ with the $i$\textsuperscript{th} constraint in \eqref{feas}, we construct the following Lagrangian function:
\begin{eqnarray} \label{lagrangian}
L(\bm{p}, \lambda, \bm{\nu}) = \sum_i \big(p_i R_{t-1, i} - \frac{1}{\eta}p_i \ln p_i - \lambda p_i -\nu_i p_i \big) 
\end{eqnarray}
For a set of dual variables $(\lambda, \bm{\nu})$, we set the gradient of $L$ w.r.t. the primal variables $\bm{p}$ to zero to obtain:
\begin{eqnarray*}
p_i &=& \exp(\eta R_{t-1,i})	\exp(\lambda \eta - \eta \nu_i -1 )\\
&=& K \exp(\eta R_{t-1,i})\zeta_i,
\end{eqnarray*}
where $K \equiv \exp(\lambda \eta -1) \geq 0$ and $\zeta_i \equiv \exp(-\eta \nu_i) \leq 1.$ Let us fix the constant $K$. To ensure the complementary slackness condition corresponding to the constraint \eqref{feas}, we choose the dual variable $\nu_i \geq 0$ such that $p_i = \min (1, K \exp(\eta R_{t-1,i})), \forall i.$ Finally, we determine the constant $K$ from the equality constraint \eqref{cap}: 
\begin{eqnarray} \label{constK}
\sum_{i=1}^N \min (1, K \exp(\eta R_{t-1,i}))=k.	
\end{eqnarray}
For any $k<N$, we now argue that the equation \eqref{constK} has a unique solution for $K>0$. The LHS of the equation \eqref{constK} is a continuous, non-decreasing function of $K$ and takes value in the interval $[0, N]$. Hence, by the intermediate value theorem, the equation \eqref{constK} has at least one solution. Furthermore, at the equality, at least one of the constituent terms will be strictly smaller than one. Since this term is strictly increasing with $K$, the proposition follows. \\
To efficiently solve the equation \eqref{constK}, we sort the cumulative request vector $\bm{R}_{t-1}$ in non-increasing order. Let $R_{t-1, (i)}$ denote the $i$\textsuperscript{th} term of the sorted vector. Let $i^*$ be the largest index for which $K \exp(\eta R_{t-1, (i^*)}) \geq 1.$
Then, the equation \eqref{constK} can be written as:
\begin{eqnarray*}
i^* + K \sum_{j=i^*+1}^N \exp(\eta R_{t-1, (j)}) = k.	
\end{eqnarray*}
\emph{i.e.,}
\begin{eqnarray} \label{Keqn}
K = \frac{k-i^*}{\sum_{j=i^*+1}^N \exp(\eta R_{t-1,(j)})}.	
\end{eqnarray}
where $i^*$ is the largest index to satisfy the following constraint:
\begin{eqnarray} \label{eq1}
	(k-i^*) \exp(\eta R_{t-1, (i^*)}) \geq \sum_{j=i^*+1}^N \exp(\eta R_{t-1,(j)}).
\end{eqnarray}
Hence, the optimal index $i^*$ may be determined in linear time by starting with $i^*=N$ and decreasing the index $i^*$ by one until the condition \eqref{eq1} is satisfied. Once the optimal $i^*$ is found, the optimal value of the constant $K$ may be obtained from equation \eqref{Keqn}. The overall complexity of the procedure is dominated by the sorting step and is equal to $O(N\ln N).$ However, since only one index changes at a time, in practice, the average computational cost is much less. 

\subsection{Approximating arbitrary Set functions by Modular functions} \label{monotone_reward}
For completeness, here we outline the main steps involved for proving Theorem \ref{thm:sandwich} (see also \cite{wu2019set} for an exposition). 
\begin{theorem}[Sandwich Theorem, \cite{iyer2012algorithms}] For a given set $X$ and any set function $f: 2^X\to\R$ and any set $Y\subseteq X$, there are two modular functions $m_u: 2^X\to\R$ and $m_l: 2^X\to\R$ such that $m_l\leq f\leq m_u$ and $m_l(Y) = f(Y)=m_u(Y)$. Furthermore, the functions $m_u$ and $m_l$ can be explicitly expressed in terms of the function $f$.

\end{theorem}
The above Sandwich theorem is a consequence of a series of results that we briefly describe below.
Recall that a set function $f: 2^{X} \to \R$ is called submodular if for all $A, B \subseteq 2^X,$ we have 
\[f(A\cup B) + f(A\cap B)\leq f(A) + f(B). \]
The following two lemmas approximate any submodular function by modular functions. For any two sets $A, B \subset 2^X$, define $f(A|B):=f(A\cup B)-f(B).$
\begin{lemma}[Upper bound \citep{iyer2012algorithms}]\label{upper}
For any submodular function $f:2^X\to \R$, and $Y\subseteq X$, there exists a modular function $m_u(A)$ such that $m_u\geq f$ and $m_u(Y)=f(Y)$. One such candidate modular function $m_u$ is given as follows:
\begin{align}
m_u(A) = f(Y) + \sum_{j \in A\setminus Y}f(j|\emptyset) - \sum_{j \in Y\setminus A} f(j|Y\setminus j).  
\end{align}
\end{lemma}

\begin{lemma}[Lower bound \citep{iyer2012algorithms}]\label{lower}
For any submodular function $f:2^X\to \R$, and $Y\subseteq X$, there exists a modular function $m_l(A)$ such that $m_l\leq f$ and $m_l(Y)=f(Y)$. One such candidate modular function $m_l$ is given as follows: \\
Define any permutation (ordering) of the elements of $X=\{x_1, x_2, \dots, x_{|X|}\}$. Subsequently define $Y=\{x_1, x_2, \dots, x_{|Y|}\}$ and sets $S_i = \{x_1, x_2,\dots, x_i\}$. Define $m_l(\emptyset)=f(\emptyset)$. Then, for $\emptyset\neq A\subseteq X$,
\begin{align}
    m_l(A) = m_l(\emptyset) + \sum_{x_i\in A}(f(S_i)-f(S_{i-1})).
\end{align}
\end{lemma}
Finally, the following result shows that any arbitrary set function can be expressed as the difference of two submodular functions.
\begin{lemma}[Difference of Submodular functions \citep{narasimhan2012submodular}]\label{dsfunc}
Every set function $f:2^X\to \R$ can be expressed as the difference of two monotone nondecreasing submodular functions $g$ and $h$, i.e., $f=g-h$.
\end{lemma}

\cite{iyer2012algorithms} gives an exact characterization of the functions $g$ and $h$ as follows: let $h$ be any strictly submodular function. Compute
\begin{align}
    \beta = \min_{Y\subset Z\subseteq X\setminus j} \bigg(h(j|Y)-h(j|Z)\bigg).
\end{align}
For example, by taking $h(Y):=\sqrt{|Y|},$ we have $\beta=2\sqrt{N-1}-\sqrt{N}-\sqrt{N-2}=O(\frac{1}{N^{3/2}}),$ where $N=|X|$. Similarly, define
\begin{align}
    \alpha(f) = \min_{Y\subset Z\subseteq X\setminus j} \bigg(f(j|Y)-f(j|Z)\bigg).
\end{align}
By definition $\alpha\geq0\iff f$ is submodular. In that case, we can take $g=f, h=0$ and we are done. 
In case $\alpha<0,$ consider any $\alpha'\leq \alpha$. Then, \cite{iyer2012algorithms} showed that the function $f$ can be expressed as $f = \hat{g}-\hat{h}$ where
\begin{align}
    \hat{g} = f + \frac{|\alpha'|}{\beta}h, ~\text{and}~
    \hat{h} = \frac{|\alpha'|}{\beta}h,
\end{align}
where $\hat{g}$ and $\hat{h}$ can be easily seen to be submodular. 

Note that computing the parameter $\alpha(f)$ for any arbitrary set function $f$ could be intractable~\citep{iyer2012algorithms}. However, we can readily obtain a lower bound $\alpha'$ to $\alpha$ for monotone reward functions. We have
\begin{align*}
    \alpha &= \min_{Y\subset Z\subseteq X\setminus j} f(j|Y)-f(j|Z)\\
    &\geq \min_{Y\subseteq X}f(j|Y)-\max_{Z\subseteq X}f(j|Z)\\
    &\stackrel{(a)}{\geq} -\max_{Z\subseteq X}f(j|Z)=:\alpha',
\end{align*}
where the inequality (a) follows from the monotonicity of the function $f$.
In other words, $|\alpha'|$ is largest marginal gain of adding an element to any set $Z\subseteq X$ for the function $f$. To proceed further, we assume the function $f$ to be smooth, \emph{i.e.,} $\forall S\subseteq [N], x\in S,$ one has: 
\begin{eqnarray} \label{smoothness_assumption}
|f(S)-f(S\setminus \{x\})|\leq G,
\end{eqnarray}
for some finite constant $G.$
Under the smoothness assumption, we can set $|\alpha'|=G.$ Combining \cref{upper}, \cref{lower}, and \cref{dsfunc}, we can now explicitly write down the expressions for the modular functions $m_l$ and $m_u$ appearing on \cref{thm:sandwich} as follows:
\begin{align}
     m_l &= m_l^g - m_u^h\\
     m_u &= m_u^g - m_l^h.
\end{align}
In the following, we derive an explicit expression for each components of the modular function $m_l$.

\textbf{Expression for the function $m_l$:}
For a given ordering $\pi$ of the elements, let $\sigma(i) \equiv \pi^{-1}(i)$ denote the position of the element $i$ in the ordering. 
Setting $Y=[N]$ and choosing $h(S)= \sqrt{|S|}$ in \cref{upper} and \cref{lower}, we have the following expression for the function $m_l$:
\begin{align*}
    m_l^{\hat{g}}(i)&=\hat{g}(S_{\sigma(i)}) - \hat{g}(S_{\sigma(i)-1})\\
     &=f(S_{\sigma(i)}) - f(S_{\sigma(i)-1}) + \frac{|\alpha'|}{\beta}\left(h(S_{\sigma(i)})-h(S_{\sigma(i)-1})\right)\\
    &=f(S_{\sigma(i)}) - f(S_{\sigma(i)-1}) + \frac{|\alpha'|}{\beta}\left(\sqrt{|S_{\sigma(i)}|}-\sqrt{|S_{\sigma(i)-1}|}\right)
    \end{align*}
    Furthermore, we have
    \begin{align*}
    m_u^{\hat{h}}(i) &=\frac{|\alpha'|}{\beta}\left(h([N]) + \sum_{j \in i\setminus [N]}h(j|\emptyset) - \sum_{j \in [N]\setminus i} h(j|[N]\setminus j)  \right)\\
    &=\frac{|\alpha'|}{\beta}\left(h([N]) - \sum_{j \in [N]\setminus i} h(j|[N]\setminus j)  \right)\\
    &=\frac{|\alpha'|}{\beta}\left(\sqrt{N} - (N-1)(\sqrt{N}-\sqrt{N-1}) \right)=:C
\end{align*}
Hence, the $i$\textsuperscript{th} component of the function $m_l$ is given by:
\begin{eqnarray} \label{g-vals}
    g(i) &\equiv& m_l(i)\nonumber \\
    &=&m_l^{\hat{g}}(i)-m_u^{\hat{h}}(i) \nonumber \\
    &=& f(S_{\sigma(i)}) - f(S_{\sigma(i)-1}) + \frac{|\alpha'|}{\beta}\left(\sqrt{|S_{\sigma(i)}|}-\sqrt{|S_{\sigma(i)-1}|}\right)-C.
\end{eqnarray}
Eqn.\ \eqref{g-vals} gives an explicit and efficiently computable expression for the lower modular function $m_l$ which we use in our online learning policy. 
Now notice that
\begin{align*}
   \sqrt{N} - (N-1)(\sqrt{N}-\sqrt{N-1}) &= \sqrt{N}\left(1-(N-1)\left(1-\sqrt{1-\inv{N}}\right)\right) \\
   &\approx \sqrt{N}\left(1-\frac{N-1}{2N}\right)=O(\sqrt{N})
\end{align*}
Hence, $C=\frac{|\alpha'|}{\beta}(\sqrt{n} - (N-1)(\sqrt{N}-\sqrt{N-1}))=O(FN^2)$
Using triangle inequality,
\begin{align} \label{magnitude-bound}
    |g(i)| &\leq |f(S_{\sigma(i)}) - f(S_{\sigma(i)-1})| + \frac{|\alpha'|}{\beta}\left|\sqrt{|S_{\sigma(i)}|}-\sqrt{|S_{\sigma(i)-1}|}\right|+|C|\nonumber \\
    &\stackrel{(a)}{\leq} G + \frac{G}{\beta} + |C|=O(GN^2)
\end{align}
where (a) holds because by assumption $f$ is smooth with parameter $G, \beta \sim O(1/N^{3/2}),$ $|\alpha'| \leq G.$  
Hence, the sum of the largest $k$ components of the vector $(g^2_1, g^2_2, \ldots, g^2_N)$ can be bounded by: 
\begin{eqnarray*}
	||\bm{g}^2||_{k, \infty} \leq O((\sqrt{k}GN^2)^2). 
\end{eqnarray*}
Note that the upper bound in \cref{magnitude-bound} holds for any set function $f.$ As shown below, the above bound can be improved in the special case when the function $f$ is known to be submodular.   
\paragraph{Special Case - Submodular $f:$ }
As discussed above, if the function $f$ is restricted to be submodular, we can directly use \cref{lower} to obtain an expression for the modular function $m_l$ as follows: Fix any permutation of the elements of $[N].$
\begin{eqnarray*}
	g(i) = m_l(i) = f(S_{\sigma(i)}) - f(S_{\sigma(i)-1}). 
\end{eqnarray*}
This gives the following bound $|g(i)|\leq G, \forall i\in [N].$ Hence, proceeding as above, we have:
\begin{eqnarray*}
||\bm{g}^2||_{k, \infty}\leq O((\sqrt{k}G)^2).	
\end{eqnarray*}

\subsection{Proof of Theorem \ref{k-experts-lower-bound}}\label{k-experts-lower-bound-proof}
\textbf{Outline:} We seek to obtain a tight lower bound to the regret of the $k$-\texttt{experts} problem with the \texttt{Max}-reward variant. Before we delve into the technical details, we first outline the main steps behind the proof. We define an i.i.d. reward structure where the reward of any expert at each slot is distributed as i.i.d. Bernoulli with parameter $p = \nicefrac{1}{2k}.$ Next, we compute a lower bound to the expected cumulative reward accrued by the static offline oracle policy by constructing a set $S^*$ consisting of $k$ experts, as outlined next. First, we divide the set of $N$ experts into $k$ disjoint partitions, each consisting of $\frac{N}{k}$ experts \footnote{For ease of typing, we assume that the number of experts $N$ is divisible by $k.$ If that is not the case, consider the first $\tilde{N}= k \lfloor \frac{N}{k} \rfloor$ experts only.}. Denote the set of experts in the $i$\textsuperscript{th} partition by $P_i, 1\leq i \leq k.$ Let $e^*_i \in P_i$ be the expert from the $i$\textsuperscript{th} having the highest cumulative reward up to time $T$ in hindsight. Finally, we define the set $S^* \equiv \{ e^*_i, 1\leq i \leq k\}.$ Trivially, the cumulative reward accrued by the optimal offline oracle is lower bounded the reward accrued by the set of experts in $S^*$. Furthermore, since the experts $e^*_i, 1\leq i \leq k $ are identically distributed and independent of each other, the computation of the reward accrued by the set $S^*$ becomes tractable. In the following, we show that the expected reward accumulated by the set $S^*$ is given by the expectation of the maximum of $k$ i.i.d. Binomial random variables. The regret lower bound in Theorem \ref{k-experts-lower-bound} finally follows from a tight non-asymptotic lower bound to this expectation, which we believe, has not appeared in this form before.
\begin{proof}
We use the standard ``randomization trick'' to obtain a lower bound to the worst-case regret:
\begin{align} 
    \max_{\{\bm{r}_t\}_{t=1}^T} \mathcal{R}_T & \ge \mathbb{E}_r\big(\mathcal{R}_T\big),
\end{align}
where we use the symbol $\mathbb{E}_{r}$ to convey that the expectation is taken over a random binary input reward sequence $\{r_{t,i}\}_{i\in[N], 1\le t\le T}$, where the random rewards $r_{t,i}$'s are taken to be i.i.d. $\sim \mathrm{Bern}(p)$, for some parameter $p\in [0,1]$, that will be fixed later.
Using the definition of the regret in Eq.~\eqref{regret-def}, we obtain:
\begin{align} \label{regd1}
    \max_{\{\bm{r}_t\}_{t=1}^T} \mathcal{R}_T \ge \mathtt{OPT} - \sum_{t=1}^T \mathbb{E}_{r}\big({\max_{i\in S_t}r_{t,i}}\big),
\end{align}
where we denote \begin{align}
    \mathtt{OPT} & = \mathbb{E}_r \big(\max_{S\subset [N]:\abs{S}=k}\sum_{t=1}^T \max_{i\in S}r_{t,i}\big).
\end{align}
Since the rewards $r_{t,i},\ i\in [N]$ are i.i.d.$\sim \mathrm{Bern}(p)$, for any choice of the set $S_t,$ we have: 
\begin{align} \label{regd2}
    \mathbb{E}_r\big(\max_{i\in S_t}r_{t,i}\big) = \mathbb{P}\big(\max_{i\in S_t}r_{t,i} = 1\big) = 1 - (1 - p)^k.
\end{align}
It now remains to establish a lower bound to the quantity $\mathtt{OPT}$. In order to do that, we first make the trivial observation that, for any subset $S\subseteq [N]$ with cardinality $k$, the following holds true:
\begin{align} \label{optlb}
    \mathtt{OPT} & \ge \sum_{t=1}^T \mathbb{E}\big(\max_{i\in S}r_{t,i}\big).
\end{align}
Note that in the above, we can allow random $S$, that might depend on the particular realizations of the random reward sequence. Using this observation, we now use the bound \eqref{optlb} with the set $S^\star$ as defined below:
Divide the set $N$ experts into $k$ disjoint partitions $B_1,\cdots, B_k$, each of size $b =  N / k$, such that 
\begin{align}
\label{eq:super-bins-1-to-k-1}
B_l & = \{(l-1)b+1,\cdots, lb\},\ ~1\le l\le k.
\end{align}
Finally, we construct the set $S^\star \equiv \{i_1,\cdots, i_k\}$, where, $i_l = \arg\max_{j\in B_l}X_{T,j},\ 1\le l \le k$, where $X_{T,j}=\sum_{t=1}^T r_{t,j}$. In other words, $i_l$ is the (random) index of the expert in the $l$\textsuperscript{th} partition such that it has the highest cumulative reward in hindsight. By construction, the random indices $i_1,\cdots i_k$ are independent of each other. Hence, the random rewards $r_{t,i},\ i\in S^\star$ are independent Bernoulli random variables with some parameter $q$, that we will determine shortly. Using the observation that for a fixed $1\le l\le k$, the random variables $r_{t,i_l}$ for $t=1,\cdots, T$, are identically distributed, it follows that $\mathbb{E}(r_{t,i_l})$ is identical for all $t$ for a fixed $l$, so that
\begin{align}
    q \equiv \mathbb{E}(r_{t,i_l}) = \frac{1}{T}\mathbb{E}(X_{T,i_l})= \frac{1}{T}\mathbb{E}(\max_{j\in B_l}X_{T,j}).
\end{align}
Hence, using the lower bound \eqref{optlb}, we have
\begin{align} \label{regd3}
    \mathtt{OPT} & \ge \sum_{t=1}^T \big(1 - (1 - q)^{k}\big)= T(1-(1-q)^k).
\end{align}
Hence, combining Eqns.\ \eqref{regd1}, \eqref{regd2} with the lower bound in Eqn.\ \eqref{regd3}, we have the following regret lower bound in terms of the yet undetermined parameter $q$:
\begin{align}
  \max_{\{\bm{r}_t\}_{t=1}^T}\mathcal{R}_T & \ge T\left((1-p)^k - (1-q)^k\right).
\end{align}
Since the function $(1-p)^k$ is convex in $p$, linearizing the function around the point $q$ yields the following lower bound for regret:
\begin{align}
\label{eq:k-regret-prelim-lower-bound}
    \max_{\{\bm{r}_t\}_{t=1}^T}\mathcal{R}_T & \ge kT(q-p)(1-q)^{k-1}.
\end{align}
To proceed further, we need to estimate $q$ by finding tight upper and lower bounds for it.

\paragraph{1. Upper bounding $q$:}
\label{sec:upper-bound-q}
Since the random variables $X_{T,j},\ j\in B_1$ are i.i.d. Binomial, and hence subGaussian with mean $\mu=\mathbb{E}X_{T,1} = Tp$ and variance $\sigma^2 = Tp(1-p)$, it follows from Massart's maximal lemma for Gaussians \citep{massart2007concentration} that:
\begin{align*}
    q - p &= \frac{1}{T}\bigg( \mathbb{E}(\max_{j \in B_1} X_{T,j})-pT\bigg) \nonumber\\
     &\le \sqrt{\frac{2p(1-p)\ln(N/k)}{T}}. 
\end{align*}
In particular, for a large enough horizon-length $T \geq  8 (\frac{1}{p}-1) \ln(\frac{N}{k})$, from the above we have the following upper bound for $q$: 
\begin{eqnarray} \label{eq:q-upper-bound}
	q \leq \frac{3p}{2}.
\end{eqnarray}

\paragraph{2. Lower bounding $q$:}
\label{sec:lower-bound-q}
%
We have 
\begin{align} \label{q-lb}
    q - p & = \frac{1}{T}\mathbb{E}\bigg(\max_{j\in B_1}(X_{T,j} - Tp)\bigg)\nonumber\\
    \ & = \frac{1}{T}\mathbb{E}\bigg(\max_{j\in B_1}(X_{T,j} - Tp)\mathds{1}\big(\max_{j \in B_1}X_{T,j}<Tp\big)\bigg)\nonumber\\
    \ & + \frac{1}{T}\mathbb{E}\bigg(\max_{j\in B_1}(X_{T,j} - Tp)\mathds{1}\big(\max_{j \in B_1}X_{T,j} \geq Tp\big)\bigg)\nonumber\\
    \ & \stackrel{\textrm{(def.)}}{=} \frac{I_1 + I_2}{T}.
\end{align}
Now, we separately lower bound each of the quantities $I_1$ and $I_2$ as defined above. 
\paragraph{2.1. Lower bounding $I_1$:}
%
%
We have the following inequalities:
\begin{align} \label{I1-bd}
I_1 &\equiv \mathbb{E}\bigg( \max_{j \in B_1}(X_{T,j}-Tp) \mathds{1}\big(\max_{j \in B_1}X_{T,j} <Tp \big) \bigg)  \nonumber \\
& \stackrel{(a)}{\geq} \max_{j \in B_1} \mathbb{E}\bigg((X_{T,j}-Tp) \mathds{1}(X_{T,j} <Tp ) \prod_{i \in B_1, i \neq j}\mathds{1}(X_{T,i} <Tp)\bigg) \nonumber \\
& \stackrel{(b)}{=} \mathbb{E}\bigg((X_{T,1}-Tp) \mathds{1}(X_{T,1} <Tp )\bigg) \bigg(\mathbb{P}(X_{T,1}<Tp)\bigg)^{b-1}\nonumber \\
& \stackrel{(c)}{\geq} - \mathbb{E}\big|X_{T,1}-Tp\big| \bigg(\mathbb{P}(X_{T,1}<Tp)\bigg)^{b-1} \nonumber \\
& \stackrel{(d)}{\geq} - \sqrt{Tp(1-p)} \bigg(\mathbb{P}(X_{T,1}<Tp)\bigg)^{b-1} \nonumber \\
& \stackrel{(e)}{\geq} - \big(\frac{3}{4}\big)^{b-1}\sqrt{Tp(1-p)}.  
\end{align}
in the above,
\begin{enumerate}
\item  inequality (a) follows from Jensen's inequality and the trivial fact that $\mathds{1}(\max_{j \in B_1}X_{T,j} <Tp )= \mathds{1}(X_{T,j} <Tp ) \prod_{i \in B_1, i \neq j}\mathds{1}(X_{T,i} <Tp)$
\item  inequality (b) follows from the fact that the collection of r.v.s $\{X_{T,j}, j \in B_1\}$ are independent and identically distributed
\item  inequality (c) follows from the fact that $(X_{T,1}-Tp) \mathds{1}(X_{T,1} <Tp ) \geq - \big|X_{T,1}-Tp\big|,$ 
\item in inequality (d), we have used Jensen's inequality with the fact that $X_{T,1} \sim \textrm{Binomial}(T, p)$
\item finally, in inequality (e), we have used Theorem 1 from \cite{greenberg2014tight} which states that for $p > \nicefrac{1}{T}$ we have $\mathbb{P}(X_{T,1} \geq Tp) \ge \nicefrac{1}{4}.$ 
\end{enumerate}

\paragraph{2.2. Lower bounding $I_2$:}
Using Markov's inequality, we have for any $s\geq 0:$
\begin{align} \label{eq:I_2-lower-bound}
    I_2\ & \ge s \mathbb{P}\big(\max_{ j\in B_1}X_{T,j}>s+Tp\big)\nonumber\\
    \ & \stackrel{(a)}{=}s\left(1 - \big(\mathbb{P}\big(X_{T,1}\le s+ Tp\big)\big)^b\right) \nonumber \\
    &  \stackrel{(b)}{\geq} s \bigg(1-\big(\Phi(\frac{s}{\sqrt{Tp(1-p)}})\big)^b \bigg). 
\end{align}
where in step (a), we have used the independence of the r.v.s $X_{T,j}, j \in B_1$ and in step (b), we have used Slud's inequality \citep{cesa2006prediction}. Note that in the above, we use the standard notation where $\Phi(\cdot)$ denotes the CDF of the standard Normal variable. 

Observe that for any $u>0$, we can upper bound the normal CDF as: 
\begin{align}
    \Phi(u) & = 1 - \frac{1}{\sqrt{2\pi}}\int_u^\infty e^{-x^2/2}dx   \nonumber \\
    & \le 1 - \frac{1}{\sqrt{2\pi}}\int_u^{2u} e^{-x^2/2}dx \nonumber \\
    & \le 1-\frac{u e^{-2u^2}}{\sqrt{2\pi}}.
\end{align}
By making a change of variable $u \gets \frac{s}{\sqrt{Tp(1-p)}}$ in Eqn. \eqref{eq:I_2-lower-bound}, the quantity $I_2$ can be lower bounded as:
\begin{eqnarray} \label{I2-bd2}
I_2 \geq \sqrt{Tp(1-p)} \bigg[ u \bigg(1- \big(1-\frac{u e^{-2u^2}}{\sqrt{2\pi}}\big)^b \bigg)\bigg]. 	
\end{eqnarray}
Choosing $u=\sqrt{\frac{\ln b}{2}}$ and using the standard inequality $1-x \leq e^{-x}, \forall x$, from the above we have:
\begin{align} \label{I2-lb-final}
I_2 \geq c_1\sqrt{Tp(1-p) \ln b},	
\end{align}
 where $c_1 \equiv \frac{1}{\sqrt{2}}(1- e^{-\sqrt{\ln b/4\pi}}).$ 

Combining the bounds for $I_1$ and $I_2$ from \eqref{I1-bd} and \eqref{I2-lb-final}, we obtain the following lower bound for $q$ from Eqn.\ \eqref{q-lb} valid for $b \equiv \frac{N}{k} \geq 7$:
\begin{align} \label{q-lb-final}
	q-p \geq \frac{c_2}{T}\sqrt{Tp(1-p) \ln \frac{N}{k}},  
\end{align}
where $c_2 \geq 0.1$ is an absolute constant.
\paragraph{3. Lower bounding the regret:}
Finally, we choose $p = \frac{1}{2k}.$
Substituting the bounds \eqref{eq:q-upper-bound} and \eqref{q-lb-final} into the regret lower  bound~\eqref{eq:k-regret-prelim-lower-bound}, for $T \geq  16k \ln(\frac{N}{k})$ and $\frac{N}{k} \geq 7,$ we obtain: 
\begin{align}
    & \max_{\{\bm{r}_t\}_{t=1}^T}\mathcal{R}_T  \ge c_2 k \sqrt{\frac{T}{2k}(1-\frac{1}{2k}) \ln \frac{N}{k}} \bigg(1-\frac{3}{4k}\bigg)^{k-1} \geq c_3 \sqrt{kT \ln \frac{N}{k}},
      \end{align}
      where $c_3 \geq 0.02 $ is an absolute constant.
  \end{proof}
 \section{Additional Experimental Results} \label{addl-expts}
  \begin{figure*}[!htb]
\minipage{0.33\textwidth}
  \includegraphics[width=\linewidth]{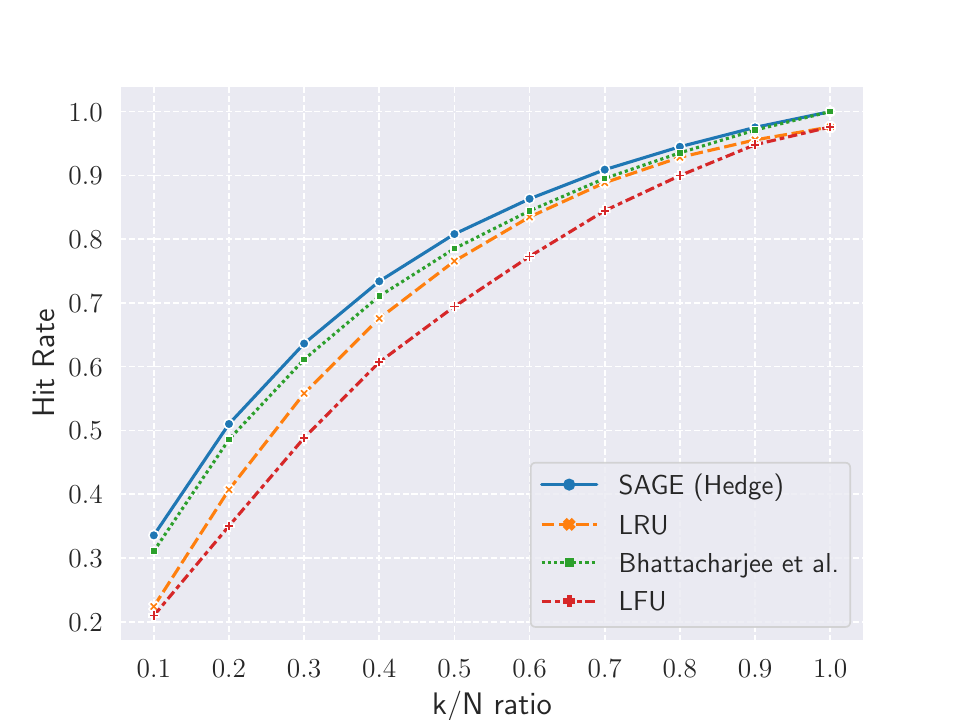}
  \caption{\small{Comparison among different prediction policies in terms of hit rates (fraction of correct predictions) for different values of $\nicefrac{k}{N}, N \sim 2400$ for the MovieLens dataset.}}\label{fig:hitrates1}
\endminipage\hfill
\minipage{0.33\textwidth}
  \includegraphics[width=\linewidth]{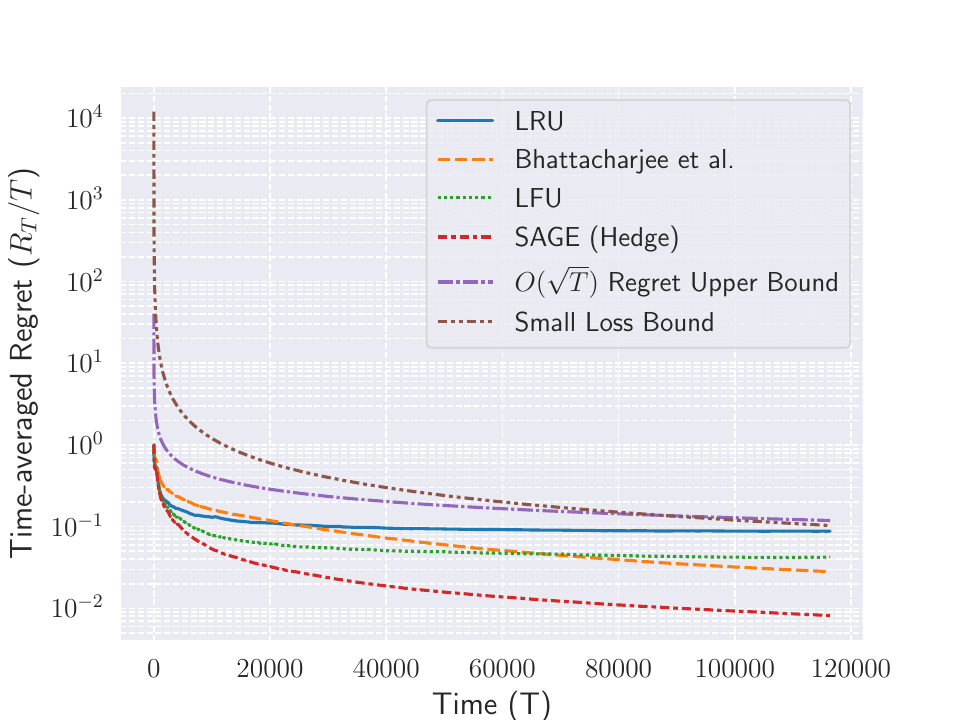}
  \caption{\small{Comparison among different prediction policies in terms of normalized regret $\frac{R_T}{T}$ with $\nicefrac{k}{N}=0.1, N \sim 2500$ for the Wiki-CDN dataset.}}\label{fig:regret2}
\endminipage\hfill
\minipage{0.33\textwidth}%
  \includegraphics[width=\linewidth]{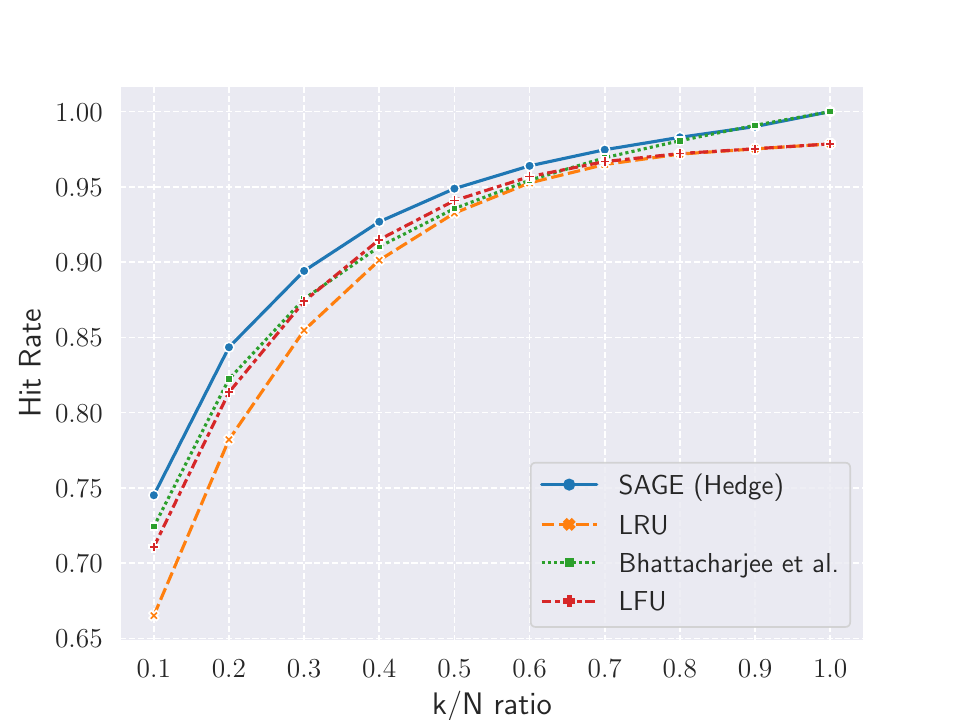}
  \caption{\small{Comparison among different prediction policies in terms of hit rates (fraction of correct predictions) for different values of $\nicefrac{k}{N}, N \sim 2500$ for the Wiki-CDN dataset.}}\label{fig:hitrates2}
\endminipage
\end{figure*}
In Figure \ref{fig:hitrates1}, we plot the hit rates (\emph{i.e.,} the fraction of correct predictions) of various prediction policies for the $\texttt{k-sets}$ problem for the MovieLens dataset. From the plots, we observe that by selecting only $30\%$ of the elements (\emph{i.e.,} $\nicefrac{k}{N}=0.3$), the $\texttt{SAGE}$ policy with $\pi_{\text{base}}=\texttt{Hedge}$ achieves a hit rate of at least $60\%$. We also measure the performance of the proposed policy for the $\texttt{k-sets}$ problem on Wiki-CDN dataset~\citep{berger2018practical}. This dataset contains publicly available Wikipedia CDN request traces from a server located in San Francisco. It contains trace for $T\sim 10^5$ time stamps and $N\sim2500$ files. We compare the performance of different policies in terms of the normalized regret and hit rates in Figure~\ref{fig:regret2} and \ref{fig:hitrates2} respectively. From the plots, we observe that the \texttt{SAGE} policy outperforms other benchmarks by a large margin.
 
\subsection{Experiments with the learning policy for Monotone Reward Functions (Section \ref{general_reward})}

\begin{figure*}[!htb]
\minipage{0.48\textwidth}
  \includegraphics[width=\linewidth]{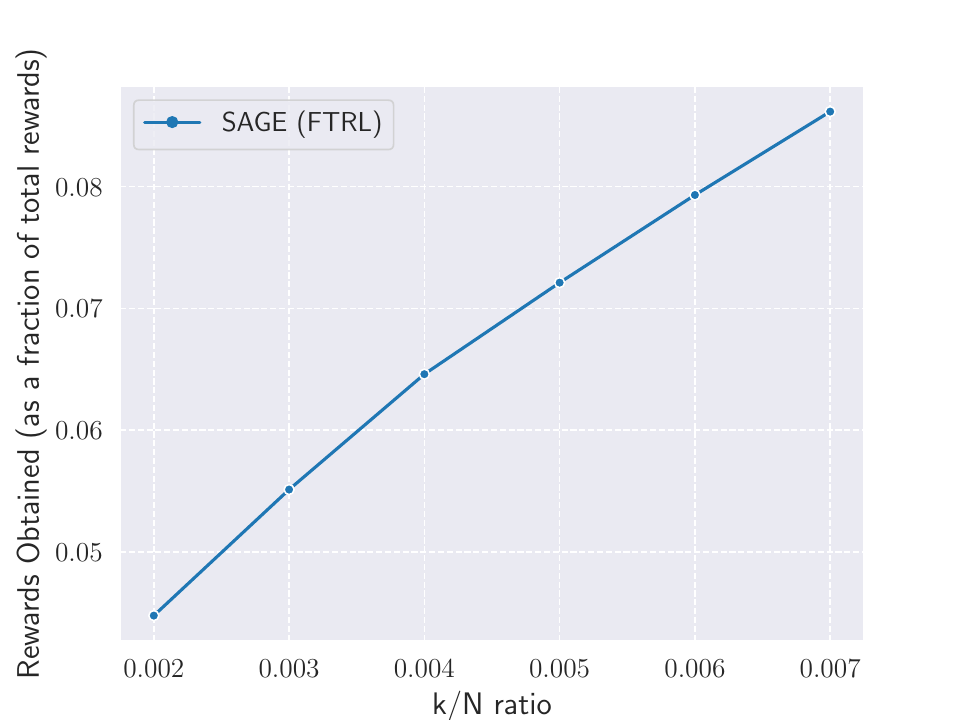}
  \caption{\small{Performance of prediction policy in terms of fraction of total possible reward (by selecting all the elements) obtained for $N=1000,  v:\bm{x}\mapsto ||\bm{x}||_2$ for the MovieLens dataset.}}\label{fig:p_2_reward_frac}
\endminipage\hfill
\minipage{0.48\textwidth}%
  \includegraphics[width=\linewidth]{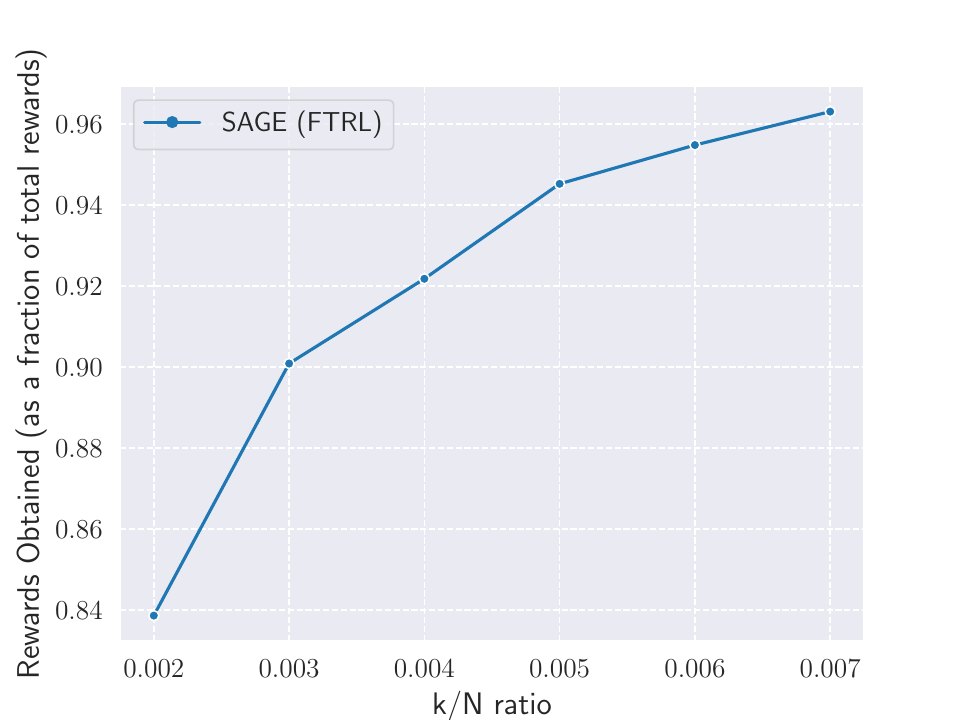}
  \caption{\small{Performance of prediction policy in terms of fraction of total possible reward (by selecting all the elements) obtained for $N=1000,  v:\bm{x}\mapsto ||\bm{x}||_\infty$ for the MovieLens dataset.}}\label{fig:p_inf_reward_frac}
\endminipage
\end{figure*}

In our experiments for the general monotone reward functions, we use a subset of the MovieLens dataset with $T\sim 200$ and $N=100$. Similar to Section \ref{sims}, we assume that the movies are sorted according to genres so that if movie $i$ is chosen by the user at round $t$, then the reward vector, $\bm{r}_t\in[0,1]^N$, is given as $r_{t,j}=1-\frac{1}{N}|j-i|$. For a reward vector $\bm{r}_t$ and real-valued function $v:\mathbb{R}^N\to\mathbb{R}_{\geq0}$, we define a monotone set function $f_t:2^{[N]}\to\mathbb{R}_{\geq0}$ as $f_t(S) = v(\bm{r}_t(S)), \forall S\subseteq [N]$ where $[r_t(S)]_i=r_{t,i}\cdot\mathds{1}(i\in S)$. According to the \texttt{$k$-experts} setting, we assume that the learner receives a reward of $f_t(S_t)$ for predicting the set $S_t$. In our experiments, we consider two different reward functions $v:\bm{x}\mapsto ||\bm{x}||_p$, with $p=2$ and $p=\infty$. 
In Figure~\ref{fig:p_2_reward_frac} and~\ref{fig:p_inf_reward_frac}, we plot the rewards obtained by the learner as a fraction of the total possible rewards (when all the elements are selected). From the plots, it is clear that the proposed policy has excellent performance for both reward functions, as it achieves a large fraction of the total possible reward by using only a small fraction of the experts.   

\end{document}